\documentclass[onecolumn,usenatbib,amsmath,amssymb,amsbsy]{mn2e}

\usepackage[T1]{fontenc}
\usepackage{aecompl}

\usepackage{natbib}
\usepackage{amsbsy}
\usepackage{amssymb}
\usepackage{amsmath}
\usepackage[dvipsnames,usenames]{color}
\usepackage{soul}

\usepackage{graphicx}

\newcommand{\be}{\begin{equation}}
\newcommand{\ee}{\end{equation}}
\newcommand{\bear}{\begin{eqnarray}}
\newcommand{\eear}{\end{eqnarray}}

\newcommand{\rx}{{\rm x}}

\newcommand{\rn}{{\rm n}}
\newcommand{\rp}{{\rm p}}

\newcommand{\rnp}{{\rm np}}

\newcommand{\rP}{{\mathcal P}}

\newcommand{\rA}{{\rm A}}
\newcommand{\rE}{{\rm E}}
\newcommand{\rB}{{\rm B}}
\newcommand{\cN}{{\cal N}}

\newcommand{\cT}{{\cal T}}
\newcommand{\cB}{{\cal B}}

\newcommand{\cI}{{\cal I}}

\begin{document}

\title[Crust-core spin lag in neutron stars] 
{Persistent crust-core spin lag in neutron stars}

\author[Glampedakis \& Lasky]{Kostas Glampedakis$^{1,2}$ and Paul D. Lasky$^{3,4}$ \\
  \\
  $^1$ Departamento de F\'isica, Universidad de Murcia, Murcia, E-30100, Spain \\
  $^2$ Theoretical Astrophysics, University of T\"{u}bingen, Auf der Morgenstelle 10, T\"{u}bingen, D-72076, Germany \\
 $^{3}$ Monash Centre for Astrophysics, School of Physics and Astronomy, Monash University, VIC 3800, Australia\\
 $^{4}$ School of Physics, University of Melbourne, Parkville, Victoria 3010, Australia
}

\maketitle

\begin{abstract}

It is commonly believed that the magnetic field threading a neutron star provides the ultimate mechanism (on top of fluid viscosity)   
for enforcing long-term corotation between the slowly spun down solid crust and the liquid core. We show that this argument 
fails for axisymmetric magnetic fields with closed field lines in the core, the commonly used `twisted torus' field being the most prominent 
example. The failure of such magnetic fields to enforce global crust-core corotation leads to the development of a persistent spin lag between  
the core region occupied by the closed field lines and the rest of the crust and core. We discuss the repercussions of this spin lag
for the evolution of the magnetic field, suggesting that, in order for a neutron star to settle to a stable state of crust-core corotation, the bulk 
of the toroidal field component should be deposited into the crust soon after the neutron star's birth.   

 \end{abstract}

\begin{keywords}
dense matter -- stars: neutron -- stars: magnetic field
\end{keywords}

%%%%%%%%%%%%%%%%%%%%%%%%%%%%%%%%%%%%%%%%%%%%%%%%%%%%%%%%%%%%%%%%%%
%%%%%%%%%%%%%%%%%%%%%%%%%%%%%%%%%%%%%%%%%%%%%%%%%%%%%%%%%%%%%%%%%%

\section{Introduction}
\label{sec:intro}
Conventional wisdom states that a strong magnetic field penetrating the core and crust of a neutron star enforces corotation between these two components.  It is argued that the magnetic coupling timescale is much faster than the viscous Ekman coupling timescale and the typical spindown timescale, implying the crust and core spin down in unison.  \citet{easson79} was the first to argue that this issue is far from trivial and attached certain caveats by conjecturing that the degree of crust-core coupling is largely determined by the magnetic field topology. As an example, 
he showed that crust-core corotation on secular timescales is not possible along closed lines of a purely poloidal or purely toroidal magnetic field.  

By collating some key previous results (the most relevant being the work of \citet{abney}), \citet{melatos12} recently challenged the status quo, arguing that in realistic neutron stars with {\it unmagnetized} stratified matter, buoyancy-limited Ekman flow can maintain core super-rotation for a timescale of the order of $10^3\,\mbox{yr}$.
If true, the consequences for the physics of young neutron stars (e.g. Crab-like pulsars and magnetars) are varied and rich.  For example, one might expect ongoing generation of magnetic field through a persistent dynamo, gravitational wave emission from hydrodynamic instabilities and an extra source of pulsar timing noise. 

Spherical, stratified Ekman flow {\it with} a magnetic field is currently an unsolved problem.  \citet{melatos12} argued phenomenologically that stratification lengthens 
the magnetic coupling timescale (see also~\citet{mendell98}) and concluded that for weakly magnetized systems this timescale could exceed the viscous timescale,
again allowing for persistent crust-core spin lag. Melatos also emphasised the crucial role realistic magnetic field geometry plays in determining the efficiency of magnetic coupling (irrespective of field strength), hinging his argument on the aforementioned analysis of \citet{easson79}.

It should be emphasized that the two previous arguments, namely, the buoyancy-modified magnetic coupling and Easson's crust-core
(no) corotation calculation, are distinct: the former is related to the \textit{short-term dynamical response} of the core to a change in the crust's rotational frequency,
and the latter investigates if an assumed corotating crust-core system is a consistent quasi-equilibrium state on \textit{secular spindown timescales}.  
This timescale-based distinction is self-consistent, provided the short-term magnetic coupling timescale, $t_\rB$, is much shorter that the electromagnetic 
spindown timescale, $t_{\rm sd}=\Omega/|\dot{\Omega}|$, and the viscous Ekman coupling timescale, $t_\rE$.  For a neutron star of spin period $P$ and 
typical internal magnetic field $B$  the magnetic coupling timescale is estimated to be $t_\rB \approx 5 \times 10^4\, (B/10^{12}\,\mbox{G})^{-5/3} (P/10\,\mbox{ms})^{-2/3}\,\mbox{s}$ \citep{mendell98,melatos12}, which becomes even shorter if one allows for a superconducting core: $t_\rB \approx 15\,(B/10^{12}\,\mbox{G})^{-5/6} (P/10\,\mbox{ms})^{-2/3}\,\mbox{s}$ (this is derived using the Alfv\'en velocity $v_\rA^2=H_c B_0/4\pi\rho_\rp$, where $H_c$ is the lower critical field, and $\rho_\rp$ is the energy-density of the proton fluid).  Barring non-superconducting systems with $B\ll10^{12}\,\mbox{G}$, one always has 
$t_\rB \ll {\rm min}(t_\rE, t_{\rm sd})$, implying the distinction between secular and dynamical timescales is valid.

Investigating the long-term, magnetic field-enforced, crust-core corotation is, therefore, a well defined problem and is the subject of this work.
Although Easson's (\citeyear{easson79}) analytic result is incredibly powerful, it suffers from only being applicable to a small subset of possible field configurations, all of which are known to be unstable on dynamical timescales \citep[e.g.,][]{wright73,braithwaite06a,lasky11}.  

In this paper, we fortify and extend the results of \citet{easson79} to understand this subtle, yet important problem of magnetic coupling between a neutron star's crust and core.  In doing so, we incorporate linked magnetic fields in which a toroidal component weaves through the closed-field line region of the poloidal field.  Such `twisted-torus' field configurations \citep[e.g.,][]{braithwaite06b,ciolfi09} are commonly invoked in the literature as `realistic'. 

Our main result is the following: {\it the existence of any closed poloidal-field-line region in the core of a neutron star causes that region to be magnetically decoupled from the rest of the star.}  Any region of fluid with open field lines (i.e., those that penetrate the crust-core interface) will corotate with the crust, and the two components will therefore spin down in unison.  A velocity lag will therefore build up between these open-field-line-regions and those with closed poloidal field lines.

Such a velocity lag between any two fluids is prone to magnetohydrodynamical instabilities that will undoubtably disturb the magnetic field equilibrium.  We conjecture that such instabilities cause the toroidal component of a twisted-torus-like magnetic field to evolve into the crust of the star.  The steady-state configuration has the stellar core with only open field lines, and hence the entire core corotates with the crust, while the closed poloidal field lines and toroidal field is entirely contained within the crust.  
For newly-born neutron stars with magnetar-like field strengths, evolution to this steady state can happen in the first few days to weeks of the star's life (unless if the birth field is too high, in which case the star spins down before the formation of the crust -- we quantify and discuss this below). We therefore have a natural mechanism for depositing strong toroidal fields in magnetar crusts, thereby supporting theoretical assertions that those strong magnetic field components exist (e.g., based on theoretical modelling of the observed temperatures of magnetars \citep{kaminkeretal06,HGA12}).   

The paper is organised as follows.  In Section \ref{sec:spindown} we outline our basic formalism for understanding the quasi-equilibrium during spindown and the
torque exerted between the crust and the core from the magnetic field. In Section \ref{sec:poloidal} we study crust-core corotation for the case of purely 
poloidal fields. The more realistic case of mixed poloidal-toroidal fields is discussed in Section~\ref{sec:mixed}. Section~\ref{sec:implications} is devoted to
the discussion of the astrophysical implications of an incomplete magnetic crust-core coupling. In Section~\ref{sec:conclusions} 
we summarize our results and provide pointers for future work. We recommend the `fast-track' reader (who is unwilling to navigate the mathematically 
arduous Sections \ref{sec:spindown} -- \ref{sec:mixed}) move directly to Section \ref{sec:implications}, where they will find the astrophysical implications of our results.

%%%%%%%%%%%%%%%%%%%%%%%%%%%%%%%%%%%%%%%%%%%%%%%%%%%%%%%%%%%%%%%%%%
%%%%%%%%%%%%%%%%%%%%%%%%%%%%%%%%%%%%%%%%%%%%%%%%%%%%%%%%%%%%%%%%%%

\section{Crust-core spindown}
\label{sec:spindown}

\subsection{Basic formalism}
\label{sec:formalism}

In this section, we describe the equations that govern the secular dynamics of a neutron star's fluid core during spindown. 
The solid crust is assumed to rotate rigidly with angular frequency $\mathbf{\Omega}(t)$, which is a slowly decreasing 
function of time and can be identified with the star's observed spin frequency.  Throughout the paper, we work in the instantaneous rest frame 
of the crust, in which the two fluid components of the core, i.e. the neutrons and the proton-electron conglomerate, have velocities 
$\mathbf{v}_\rn$ and $\mathbf{v}_\rp$ respectively. 
The two velocities are allowed to be distinct in order to account for the likely presence of superfluidity in the neutron and proton fluids. 
However, we shall ignore the likely superconducting nature of the protons and treat the magnetic field `classically' in terms
of the usual magnetohydrodynamic Lorentz force.

The long-term dynamics of the charged fluid in the core (protons and electrons) is described by the Euler equation\footnote{In \cite{easson79}, the Euler 
equation for the charged fluid features an extra force, $\mathbf{F}_\rB$, attributed to the electromagnetic spindown torque ` \ldots exerted directly on the plasma'. 
We believe this to be erroneous; the electromagnetic spindown torque is exerted on the stellar surface (crust) and its effect is mediated to the liquid core through 
the crust-core boundary and the induced change in $\mathbf{B}$. In other words, $\mathbf{F}_\rB$ is already accounted for in $\mathbf{F}_{\rm mag}$. 
Strangely, \citet{easson79} drops this force term before concluding his calculation, implying our calculations and Easson's converge.},  
\be
2  \mathbf{\Omega} \times \mathbf{v}_\rp + \dot{\mathbf{\Omega}} \times \mathbf{r} +   \nabla \Psi_\rp  = 
\frac{1}{\rho_\rp} \mathbf{F}_{\rm mag}  - \frac{1}{\rho_\rp} \mathbf{F}_{\rm cpl},
\label{eulerp}
\ee
where $\mathbf{F}_{\rm mag}$ is the magnetic force, $\mathbf{F}_{\rm cpl} $ is the coupling force between the neutral and charged fluids, which can include vortex mutual friction (see below). Throughout the paper, we use a subscript $\rx=\{\rp,\,\rn\}$ to describe proton and neutron fluids respectively. In general, the Euler equations feature extra terms related to the entrainment coupling between the neutron and proton fluids, however this effect is relatively weak in the core and does not change qualitatively any of the main results of this paper. We thus opt for working with a slightly `lighter' formalism with entrainment omitted. 
It should be noted that the omission of entrainment and proton superconductivity  
takes place only on the macroscopic scale of hydrodynamics while on the mesoscopic scale of individual vortices both effects are needed and are implicitly 
assumed if we are to speak of efficient, vortex-mediated, mutual friction coupling between the fluids~\citep{als84}.

The left-hand-side of (\ref{eulerp}) features the inertial Coriolis and Poincar\'e forces and the gradient of
\be
\Psi_\rp =  \tilde{\mu}_\rp  + \Phi -\frac{1}{2} | \bf{\Omega} \times \bf{r} |^2,
\label{Psi_p}
\ee
which is the combination of the proton-electron chemical potential (per unit mass) and the gravitational-centrifugal potential.

The neutron superfluid is described by a second Euler equation
\be
2  \mathbf{\Omega} \times \mathbf{v}_\rn + \dot{\mathbf{\Omega}} \times \mathbf{r} +   \nabla \Psi_\rn  = 
\frac{1}{\rho_\rn} \mathbf{F}_{\rm cpl},
\label{eulern}
\ee 
where $\Psi_\rn$ is given by an expression similar to (\ref{Psi_p}). 

In the Euler equations we have omitted the inertial acceleration terms, which are negligibly small when $\mathbf{v}_\rp$ and $\mathbf{v}_\rn$ are small. This approximation implies a \textit{nearly corotating} crust-charged fluid system. The state of exact corotation is described by $\mathbf{v}_\rp=0$, in which case the proton and neutron fluids are rigidly rotating  about the same axis as that of the crust (which we identify as the $z$-axis) with angular frequencies as measured in the inertial frame $\Omega_\rn = \Omega_\rp = \Omega$. Then, for the nearly-corotating crust-core system we can assume $\Omega_\rp, \Omega_\rn \approx \Omega$
without $\Omega_\rn$ and $\Omega_\rp$ necessarily being equal.

Equation (\ref{eulerp}) describes the departure of the system from a `background', time-independent state of exact corotation, where this departure is driven 
by $\dot{\mathbf{\Omega}}$.
As a consequence, all quantities appearing in (\ref{eulerp}) are perturbations with respect to the corotating state. 
For instance, the magnetic field can be decomposed into a fixed background part, $\mathbf{B}_0$, plus a small, spindown-induced
perturbation, $\mathbf{b} \equiv \delta \mathbf{B} $, i.e.
\be
\mathbf{B} \approx \mathbf{B}_0 + \mathbf{b}.
\ee
The magnetic force can therefore be expressed to leading order
\be
\mathbf{F}_{\rm mag} \approx  - \nabla \left ( \frac{\mathbf{B}_0 \cdot \mathbf{b} }{4\pi} \right ) + \frac{1}{4\pi} \left [\,
(\mathbf{b} \cdot \nabla ) \mathbf{B}_0 +  (\mathbf{B}_0 \cdot \nabla ) \mathbf{b} \,\right ].
\label{Fmag1}
\ee

The nature of the coupling force, $\mathbf{F}_{\rm cpl}$, is determined by the interaction between the superfluid's quantized 
vortices and the other components of the star (e.g. protons, electrons, the magnetic field). 
The simplest form for this force is provided by the Hall \& Vinen mutual friction force \citep{hall56a,hall56b}
\be
\mathbf{F}_{\rm cpl} =  2\Omega_\rn \rho_\rn \left\{\, \cB  \left[ \hat{\mathbf{z}}   \times ( \hat{\mathbf{z}}  \times \mathbf{w}) \right]+ \cB^\prime ( \hat{\mathbf{z}}  \times \mathbf{w} )  \, \right\},
\label{Fmf}
\ee
where $\mathbf{w}= \mathbf{v}_\rn -\mathbf{v}_\rp$ is the velocity lag between the neutrons and the charged particles, $\cB$ and $\cB^\prime$ are 
mutual friction coefficients  (which we take to be uniform throughout the star) and a `hat' denotes a unit vector. 
This form of the force assumes a vortex array aligned with the common spin axis.
According to our setup of the problem we have,
\be
 \mathbf{w} \approx \varpi (\Omega_\rn -\Omega_\rp ) \hat{\varphi} \equiv \varpi \Omega_\rnp \hat{\mathbf{\varphi}},
\ee
where $\varpi = r\sin\theta$ is the usual cylindrical radius. The mutual friction coupling force becomes
\be
\mathbf{F}_{\rm cpl} =  -2\varpi \rho_\rn\Omega_\rn \Omega_\rnp  \left (\, \cB \hat{\varphi} 
 + \cB^\prime \hat{\varpi} \, \right ).
 \label{Fcpl2}
\ee
The most commonly considered mutual friction mechanism in neutron star cores is the scattering of electrons by the magnetic
field of individual vortices, in which case $\cB^\prime \ll \cB \sim 10^{-4}$~\citep{als84, NA06}.

It should be noted that if (for whatever reason) superfluidity is absent, $\mathbf{F}_{\rm cpl}$ would represent the
force due to inter-particle collisions and the core would effectively behave as a single-fluid system with one common velocity, described 
by  eqn.~(\ref{eulerp}) with the appropriate readjustment of the density and pressure.  The entire analysis of Sections \ref{sec:spindown}--\ref{sec:mixed} still hold in the absence of superfluidity.

Following~\cite{easson79}, it is easy to show that the Coriolis force can also be omitted in eqns.~(\ref{eulerp}) and (\ref{eulern}).  The argument goes as follows: the Poincar\'e force in (\ref{eulerp}) is the term driving the magnetic field evolution. We estimate the strength of the magnetic field perturbation, $b$, by balancing 
the magnetic and Poincar\'e forces:
\be
F_{\rm mag} \sim \rho_\rp \dot{\Omega} R \quad \to \quad \frac{b}{B_0} \sim \frac{x_\rp}{2\pi} \left (\frac{\Omega R}{v_\rA} \right )^2
\frac{P}{t_{\rm sd}},
\ee
where $x_\rp = \rho_\rp/\rho$ is the proton fraction ($\rho=\rho_\rn+ \rho_\rp$ is the total density), $t_{\rm sd} = \Omega/|\dot{\Omega}|$ 
is the characteristic spindown timescale, $v_\rA^2 = B_0^2/4\pi \rho$ is the Alfv\'en speed, $P$ is the spin period and $R$ is the stellar radius. 

Our perturbative approach is justified as long as $b \ll B_0$ or, equivalently, when 
\be
B_0 \gg B_{\rm crit} \sim 10^{10}\,  \left (\frac{10\,\mbox{ms}}{P} \right )^{1/2}
\left ( \frac{1\,\mbox{yr}}{t_{\rm sd}} \right )^{1/2}\, \mbox{G},
\ee
where we have assumed canonical values for the stellar parameters, $R=10^6\,\mbox{cm},  \rho =10^{14}\,\mbox{g}\,\mbox{cm}^{-3}, x_\rp = 0.05$, and we have normalised $P$ and $t_{\rm sd}$ with new-born, rapidly spinning down systems in mind.
This condition is indeed expected to be satisfied by the vast majority of neutron stars. Had we used the superconducting expression 
for the Alfv\'en speed, $v_\rA^2 = H_c B_0/4\pi \rho_\rp$ (where $H_c$ is the so-called lower critical field), we would have obtained a negligibly small 
threshold for $B_{\rm crit}$ \citep{easson79}).

We can now show that $b \ll B_0$ implies a negligibly small Coriolis force as compared to the Poincar\'e force. 
From the perturbed induction equation for the magnetic field we have,
\be
\partial_t \mathbf{b} = \nabla \times (\mathbf{v}_\rp \times \mathbf{B}_0 ) 
\quad \to \quad v_\rp \sim \frac{b}{B_0} \frac{R}{t_{\rm sd}}.
\ee
The ratio of the two forces is then
\be
\frac{2\Omega v_\rp}{\dot{\Omega} R} \sim \frac{v_\rp t_{\rm sd}}{R}  \sim  \frac{b}{B_0} \ll 1,
\ee
implying we can safely omit the Coriolis force in the Euler equations (\ref{eulerp}) and (\ref{eulern}).  

%%%%%%%%%%%%%%%%%%%%%%%%%%%%%%%%%%%%%%%%%%%%%%%%%%%%%%%%%%%%%
%%%%%%%%%%%%%%%%%%%%%%%%%%%%%%%%%%%%%%%%%%%%%%%%%%%%%%%%%%%%%

\subsection{Easson's argument against corotation}
\label{sec:corotation}

If we assume the magnetic field is symmetric with respect to the spin axis, both the background and perturbation, $\mathbf{B}_0$ and $\mathbf{b}$, 
can be decomposed into poloidal and toroidal components
\be
\mathbf{B}_0 = \mathbf{B}_\rP + B^\varphi_0 \hat{\varphi},  \qquad \mathbf{b} = \mathbf{b}_\rP + b^\varphi\,\hat{\varphi}.
\ee 
The azimuthal component of (\ref{eulerp}) can be combined with equations (\ref{Fmag1}) and (\ref{Fcpl2}) to obtain
\be
 \mathbf{B}_\rP \cdot \nabla (\varpi b^\varphi) + \mathbf{b}_\rP \cdot \nabla (\varpi B^\varphi_0)
= 4\pi  \varpi^2 \left ( \, - \rho_\rp \dot{\Omega}  +  2 \cB \rho_\rn  \Omega_\rn \Omega_\rnp \, \right ).
\label{Ephi1}
\ee
which, for the case of a \textit{purely poloidal} background field, $\mathbf{B}_0 = \mathbf{B}_\rP$, becomes
\be
\mathbf{B}_\rP \cdot \nabla (\varpi b^\varphi)  = 4\pi  \varpi^2
\left ( \, - \rho_\rp \dot{\Omega}  + 2 \cB  \rho_\rn \Omega_\rn \Omega_\rnp \, \right ).
\label{Ephi2}
\ee
\citet{easson79} argued against crust-core corotation based on equation (\ref{Ephi2}). Suppose the field lines of $\mathbf{B}_\rP$ form a closed loop somewhere 
in the core.   The integral of the left-hand-side of (\ref{Ephi2}) along that loop vanishes (assuming $b^\varphi$ is not multivalued) while the integral of the right-hand-side is finite. Therefore, equation (\ref{Ephi2}) does not lead to a physically acceptable solution for $\mathbf{b}$, signalling the failure of the corotation approximation along closed poloidal 
lines. 

The assumption of a purely poloidal background field is overly restrictive, especially given that such fields are unstable on Alfv\'en timescales 
\citep[e.g.,][]{wright73,braithwaite06a,lasky11}, which is much shorter than the secular timescales we have considered heretofore.  Mixed fields that include poloidal and toroidal components, where the toroidal component threads the closed-field-line region of the poloidal field in the stellar core, provide dynamically stable configurations (e.g., \citealt{braithwaite06b,akgun13})\footnote{\citet{lander12} and \citet{mitchell15} have shown that such fields are unstable in non-superfluid barotropic stars, a result that is not applicable here.}.  Although such `twisted-torus' configurations are much studied, their effect on the rotational dynamics of a neutron star core-crust system are not known. In this paper, we calculate the effect various magnetic field configurations have on neutron star crust-core coupling, which relies on calculating the torque between these two components --- a task to which we now turn.

%%%%%%%%%%%%%%%%%%%%%%%%%%%%%%%%%%%%%%%%%%%%%%%%%%%%%%%%%%%%%
%%%%%%%%%%%%%%%%%%%%%%%%%%%%%%%%%%%%%%%%%%%%%%%%%%%%%%%%%%%%%

\subsection{The crust-core torque}

The magnetic coupling between the crust and core implies a non-zero traction (force per unit area), $\mathbf{t}$,
at the base of the crust.  The differential torque per unit area exerted on the crust by the core is
\be
\mathbf{\cN} = \mathbf{r} \times \mathbf{t}.
\ee 
The azimuthal component of the traction is responsible for a non-vanishing torque along the spin axis.  
This is built from the azimuthal and radial magnetic field components,
\be
 t^\varphi = \frac{1}{4\pi} B^\varphi B^r.
\ee
The corresponding torque per unit area is
\be
\cN^z = \frac{\varpi}{4\pi} B^\varphi B^r.\label{torque}
\ee
It should be noted that the torque is calculated at the base of the crust, at radius $r=R_{\rm c}$, 
in which case $\varpi = \varpi_{\rm c} = R_{\rm c} \sin\theta$. 

In the problem at hand, the magnetic field is perturbed as a response to the spindown.  Equation (\ref{torque}) can be expressed as
\be
\cN^z = \frac{\varpi}{4\pi} \left (\, b^\varphi B^r_0 + b^r B^\varphi_0 \, \right ),
\label{Nmixed}
\ee
where we assume that the background magnetic field can have both poloidal and toroidal components at the crust-core boundary.

The total torque exerted on the crust, $N_{\rm cc}$, is given by the surface integral of $\cN^z$ over the crust-core boundary 
(which, to a good approximation, can be assumed to be spherical):
\be
N_{\rm cc} = \int_{R_{\rm c}} dS \cN^z = \frac{1}{4\pi} \int_{R_{\rm c}} dS \varpi_{\rm c} \left (\, b^\varphi B^r_0 + b^r B^\varphi_0 \, \right ).
\label{N1}
\ee
The solution of the Euler equation provides $b^\varphi$ and $b^r$ in terms of $\dot{\Omega}$ and $\Omega_\rx$ which, upon substitution 
in (\ref{N1}), eventually leads to an equation of the form
\be
N_{\rm cc} = -\tilde{I} \dot{\Omega} + N_{\rm cpl}.
\label{N1a}
\ee
where $N_{\rm cpl}$ denotes the mutual friction coupling term and the coefficient $\tilde{I}$ plays the role of a moment of inertia (see below).

The torque, $N_{\rm cc}$, enters the equation of motion for the entire solid crust:
\be
I_{\rm cr} \dot{\Omega} = N_{\rm cc} + N_{\rm em},
\ee
where $I_{\rm cr}$ is the crustal moment of inertia  and $N_{\rm em}$ is the spindown torque due to the exterior magnetic field. 
Inserting (\ref{N1a}) and rearranging gives
\be
( I_{\rm cr} + \tilde{I} )  \dot{\Omega} = N_{\rm em} + N_{\rm cpl}.
\label{eom1}
\ee
Equation (\ref{eom1}) can be viewed as the equation of motion of the combined system comprising the crust and the fraction of the core with 
moment of inertia $\tilde{I}$. In other words, $\tilde{I}$ represents the fraction of the charged fluid in the core that is efficiently coupled to the 
crust and therefore spins down in unison with it on a secular timescale.  

In general, $\tilde{I}$ may \textit{not} coincide with the total moment of inertia, $I_\rp$, of the proton-electron core fluid.  
When $\tilde{I}\neq I_\rp$, the magnetic field cannot enforce corotation between the crust and the \textit{entire} charged core. 
In that case, a portion of the core fluid will fail to follow the spin evolution of the crust and, barring the action of some other non-magnetic crust-core 
coupling mechanism, a spin lag will develop between the uncoupled fluid and the rest of star. The rest of this paper is concerned with calculating 
$\tilde{I}$ for different magnetic field configurations, and hence determining which portions of the stellar core corotate with the crust.

%%%%%%%%%%%%%%%%%%%%%%%%%%%%%%%%%%%%%%%%%%%%%%%%%%%

\subsection{A note on symmetries}
\label{sec:symmetries}
Before moving on to the main part of the paper, i.e. the detailed calculation of the crust-core torque, $N_{\rm cc}$, and the
moment of inertia, $\tilde{I}$, we need to discuss the symmetry of the magnetic perturbation, $\mathbf{b}$, with respect to
the equatorial plane. Given the overall axisymmetry of the system, the components of $\mathbf{b}$ can be either symmetric or
anti-symmetric under reflection with respect to the equatorial plane, i.e., $z \to -z$.  An inspection of equation (\ref{Ephi1}) for the core 
spindown quasi-equilibrium reveals that the right-hand-side term is reflection-symmetric. Consequently, it only couples to those components of ${\bf b}$ 
that would make the left-hand-side terms of (\ref{Ephi1}) equatorially symmetric when combined with a given background field $\mathbf{B}_0$.

The magnetic fields considered in this work are characterized by an anti-symmetric radial component, $B_0^r(z)=-B_0^r(-z)$, and symmetric
$B_0^\theta(z)=B_0^\theta(-z)$ and $B^\varphi_0 (z)=B_0^\varphi(-z)$ components. This symmetry property of $\mathbf{B}_0$ implies that only an \textit{anti-symmetric} $b^\varphi$ and a \textit{symmetric} $b^r$ can make a contribution to the integrated torque $N_{\rm cc}$ (see equation (\ref{Nmixed})). 
It also means that, in order to calculate the torque, it suffices to integrate over just one hemisphere (e.g., the upper one).

%%%%%%%%%%%%%%%%%%%%%%%%%%%%%%%%%%%%%%%%%%%%%%%%%%%%%%%%%%%%%
%%%%%%%%%%%%%%%%%%%%%%%%%%%%%%%%%%%%%%%%%%%%%%%%%%%%%%%%%%%%%

\section{Crust-core corotation: poloidal fields}
\label{sec:poloidal}

We shall first study the problem of crust-core corotation for the case of a purely poloidal background field, i.e. $\mathbf{B}_0 = \mathbf{B}_\rP$,
\textit{without} closed lines anywhere in the core. The relevant equation for the spindown of the core is eqn.~(\ref{Ephi2}):
\be
\mathbf{B}_\rP \cdot \nabla (\varpi b^\varphi)  = 4\pi  \varpi^2
\left ( \, - \rho_\rp \dot{\Omega}  + 2 \cB  \rho_\rn \Omega_\rn \Omega_\rnp \, \right ).
\label{Ephi_pol}
\ee
Quite generally, the background field can be written in terms of an axisymmetric stream function, $\psi (r,\theta)$, as
\be
\mathbf{B}_\rP = \nabla \psi \times \nabla \varphi = \frac{1}{\varpi}  \nabla \psi \times \hat{\varphi}.
\ee
By construction, $\mathbf{B}_\rP$ is tangent to the $\psi=\mathrm{constant}$ surfaces (the introduction of the stream function 
also makes $\nabla \cdot \mathbf{B}_0 =0$ a trivial identity).  Hereafter we shall only consider the special case of a general \textit{dipole} field. The corresponding stream function
is 
\be
\psi = A f(r) \sin^2 \theta,
\ee
where $A$ is a constant amplitude and $f(r)$ is an arbitrary function (modulo the requirement of regularity at
the origin). The field components can then be expressed in spherical coordinates,
\be
B^r_0 =  2A \frac{f(r)}{r^2} \cos\theta, \qquad B^\theta_0 = -A \frac{f^\prime(r)}{r} \sin\theta,
\ee
where a prime denotes differentiation with respect to the argument. 

Inserting these in eqn.~(\ref{Ephi_pol}),
\be
2\frac{f}{r^2} \cos\theta  \partial_r T -\frac{f^\prime}{r^2} \sin\theta \partial_\theta T = \frac{4\pi }{A}
r^2 \sin^2 \theta
\left ( \, - \rho_\rp \dot{\Omega}  + 2 \cB \rho_\rn \Omega_\rn \Omega_\rnp \, \right ) ,
\label{pde1}
\ee 
where we have introduced the toroidal function
\be
T (r,\theta) \equiv \varpi b^\varphi.
\ee
The resulting partial differential equation is not, in general, amenable to a solution through separation of variables. 
In \cite{easson79} this equation is solved for $f(r) = 1/r$ combined with the extra simplification of a uniform density star.
This choice of magnetic field would be a suitable one for the exterior space but is clearly an unphysical model for
the field in the stellar interior. An example of a separable field that is also regular at $r=0$ is $f(r)\propto r^4$; however this also requires a
uniform density profile and therefore is of limited physical interest\footnote{It should be pointed out that the radial profiles $f\propto(r^2,r^4)$ 
represent the solutions of the dipolar Grad-Shafranov equation that describe the hydromagnetic equilibrium in uniform density neutron stars~\citep{ferraro54}.}.

%%%%%%%%%%%%%%%%%%%%%%%%%%%%
\subsection{The simple case of a uniform field}
The only separable solution admitted by realistic, non-uniform density neutron star models is $f(r)\propto r^2$, corresponding to the uniform 
field $\mathbf{B}_0 = B_0 \hat{z}$ with $B_0 =  2A $ and $B^r_0 = 2A\cos\theta, ~ B_0^\theta = -2A \cos\theta$. However, the 
cylindrical symmetry of the field suggests that it would be easier to study the problem in cylindrical rather than spherical coordinates. 
Indeed, the Euler equation (\ref{Ephi_pol}) in cylindrical coordinates becomes
\be
B_0 \partial_z (\varpi b^\varphi)  = 4\pi  \varpi^2 \left ( \, -  \rho_\rp\dot{\Omega}  + 2 \cB \rho_\rn \Omega_\rn \Omega_\rnp \, \right ).
\ee
This can be integrated without any difficulty (the integration constant is fixed by the requirement of an equatorially anti-symmetric $b^\varphi$),
\be
b^\varphi (\varpi, z) = \frac{4\pi \varpi}{B_0} \left (\,   -\dot{\Omega} \int_0^z dz^\prime \rho_\rp 
+ 2\cB \Omega_\rn \Omega_\rnp  \int_0^z dz^\prime \rho_\rn  \,\right ).
\label{b1}
\ee

The total crust-core torque from eqn.~(\ref{N1}) is given by 
\be
N_{\rm cc} = 2\pi R_{\rm c} \int_0^{R_{\rm c}} d\varpi \frac{\varpi^2}{(R^2_{\rm c}-\varpi^2)^{1/2}} \frac{B_0^r b^\varphi}{4\pi},
\ee
and the radial component of the background field is (in the upper hemisphere)
\be
B_0^r = B_0 \cos\theta =  B_0 \frac{z_{\rm c}}{R_{\rm c}} =  B_0 \frac{(R^2_{\rm c}-\varpi^2)^{1/2}}{R_{\rm c}},
\ee
where $ z_{\rm c}(\varpi) = \sqrt{R_{\rm c}^2 - \varpi^2} $ is the $z$-coordinate of a point on the spherical crust-core boundary. 

Given these expressions, the torque becomes
 \be
N_{\rm cc} = 4\pi  \int_0^{R_{\rm c}} d\varpi \varpi^3 \left (\,   -\dot{\Omega} \int_0^{z_{\rm c}} dz \rho_\rp 
+ 2\cB \Omega_\rn \Omega_\rnp  \int_0^{z_{\rm c}} dz \rho_\rn  \,\right ) 
=  -I_\rp  \dot{\Omega} + 2\cB  I_\rn \Omega_\rn \Omega_\rnp,
\label{Nuniform}
\ee
where
\be
I_\rx = 4\pi  \int_0^{R_{\rm c}} d\varpi \varpi  \int_0^{z_{\rm c} (\varpi)}   dz \, \rho_\rx \varpi^2,
\label{Ixcyl}
\ee
is the total moment of inertia of the $\rx =\{\rn,\rp\}$ fluid in the core. The equation of motion for the crust, eqn.~(\ref{eom1}), 
takes the following rigid-body form: 
\be
(I_{\rm cr} + I_\rp )\dot{\Omega} = N_{\rm em} + 2\cB  I_\rn \Omega_\rn \Omega_\rnp.
\label{uniformtorque}
\ee
The comparison of eqns.~(\ref{eom1}) and (\ref{uniformtorque}) allows us to conclude that the uniform dipole magnetic field, $\mathbf{B}_0 = B_0 \hat{z} $, 
leads to $\tilde{I} = I_\rp$, which means that the \textit{entire} proton-electron fluid of the core is coupled and corotates with the crust.  This result was first obtained 
by \cite{easson79}. We can show that the same property is true for the earlier case of the non-uniform field $f(r) = r^4$ in uniform density stars ---
see Appendix~\ref{sec:fr4} for a detailed discussion.

%%%%%%%%%%%%%%%%%%%%%%%%%%%%%%%%%%%%%%%%%%%%%%%%%%%%%%%%%%%%%
%%%%%%%%%%%%%%%%%%%%%%%%%%%%%%%%%%%%%%%%%%%%%%%%%%%%%%%%%%%%%

\subsection{Solving the corotation problem in magnetic coordinates}
\label{sec:poloidal_mag}

The ease with which the uniform field case was solved was a consequence of using a coordinate system adapted to the geometry of the 
field lines. This motivates us to use \textit{magnetic coordinates} for a general, axisymmetric poloidal field without closed lines,  with the expectation of a 
troubleless solution for the perturbed toroidal magnetic field $b^\varphi$. Moreover, the assumption of uniform matter is hereafter permanently abandoned,
leaving the stellar equation of state completely general. 

The magnetic coordinates comprise the stream function, $\psi$ (any other function $F(\psi)$ could be equally well used), the length along a field line, $\chi$, and 
the previously used azimuthal angle, $\varphi$. The set $\{\psi,\chi,\varphi\}$ define an orthogonal curvilinear coordinate system with a line
element
\be
dl^2 = h_\psi^2 d\psi^2 + h_\chi^2 d\chi^2 + h_\varphi^2 d\varphi^2,
\ee
where the metric functions $h_\psi, h_\chi, h_\varphi$ are all $\varphi$-independent. 
Appendix~\ref{sec:coordinates} provides the necessary mathematical supplement on magnetic coordinates. 

Expressed in the new coordinates, eqn.~(\ref{Ephi2}) for the spindown of the core charged fluids becomes,
\be
\frac{B_0}{h_\chi} \partial_\chi T = 4\pi \varpi^2 \left (\, -\rho_\rp \dot{\Omega} + 2\cB \rho_\rn \Omega_\rn \Omega_\rnp  \, \right ).
\ee
As anticipated, this can be trivially integrated:
\be
T (\psi,\chi) = -4\pi \dot{\Omega}  \int_0^\chi d\chi^\prime \frac{h_\chi}{B_0} \rho_\rp \varpi^2 
+ 8\pi \cB \Omega_\rn \Omega_\rnp \int_0^\chi d\chi^\prime \frac{h_\chi}{B_0} \rho_\rn \varpi^2,
\label{Tsol1}
\ee
where we have taken $\chi=0$ at the equatorial plane and enforced an anti-symmetric $b^\varphi$ (see Section \ref{sec:symmetries}).

For the remainder of this calculation, we need two geometric identities concerning our magnetic coordinates, the calculations for these are presented in Appendix \ref{sec:geom}.  Firstly, the surface element, $dS$, of a spherical surface in magnetic coordinates is
\be
dS = ( h_\varphi h_\psi )_{\chi_{\rm c}} \left [\, 1 + \left ( \frac{ h_\chi\partial_\psi R_{\rm c}}{h_\psi \partial_\chi R_{\rm c}} \right )^2 \, 
\right ]^{1/2}_{\chi_{\rm c}}  d\varphi d\psi,
\label{eqn:area}
\ee
where $\chi_{\rm c} = \chi_{\rm c} (R_{\rm c}, \psi)$ is the $\chi$-coordinate of a point on the sphere with radius $r=R_{\rm c}$.  
Secondly, for a general dipole field, $\psi = A f(r) \sin^2 \theta$, we have the exact identity
\be
(\hat{r} \cdot \hat{\chi}) \left [ 1 + \left ( \frac{h_\chi\partial_\psi r}{h_\psi \partial_\chi r} \right )^2 \, \right ]^{1/2} = 1.
\label{eqn:geomident}
\ee

The crust-core torque from eqn.~(\ref{N1}) can now be expressed as
\be
N_{\rm cc} = \frac{1}{4\pi} \int_{R_{\rm c}} dS \varpi_{\rm c} b^\varphi B^r_0  = \frac{1}{4\pi} \int_{R_{\rm c}} d\varphi d\psi \,  
\left [ 1 + \left ( \frac{h_\chi\partial_\psi R_{\rm c}}{h_\psi \partial_\chi R_{\rm c}} \right )^2 \, \right ]^{1/2}_{\chi_{\rm c}} 
( h_\varphi h_\psi  B_0^r  T )_{\chi_{\rm c}}.
\label{N2}
\ee
Using  $B_0^r = B_0 (\hat{r} \cdot \hat{\chi})_{\chi_{\rm c}} $, together with our earlier solution for $T$ in eqn.~(\ref{Tsol1}), $N_{\rm cc}$ becomes
\be
N_{\rm cc} = 4 \pi \int_{R_{\rm c}} d\psi\,  \left [ 1 + \left ( \frac{h_\chi\partial_\psi R_{\rm c}}{h_\psi \partial_\chi R_{\rm c}} \right )^2 \, 
\right ]^{1/2}_{\chi_{\rm c} }
[ h_\varphi h_\psi  B_0 (\hat{r} \cdot \hat{\chi})]_{\chi_{\rm c}} \left \{\, -\dot{\Omega} \int_0^{\chi_{\rm c}} d\chi \frac{h_\chi}{B_0} \rho_\rp \varpi^2 
+ 2 \cB \Omega_\rn \Omega_\rnp \int_0^{\chi_{\rm c}} d\chi \frac{h_\chi}{B_0} \rho_\rn \varpi^2 \, \right \}.
\ee
The integral can be simplified using $B_0=1/(h_\psi h_\varphi)$ (see eqn.~(\ref{Bpol_mc})), and the torque formula can be expressed in the desired compact form
\be
N_{\rm cc} = -\tilde{I}_\rp \dot{\Omega} + 2 \cB \Omega_\rn \Omega_\rnp \tilde{I}_\rn,
\ee
where the effective proton and neutron moments of inertia are defined as
\be
\tilde{I}_\rx = 4 \pi \int_{R_{\rm c}} d\psi\,   \left [ 1 + \left ( \frac{h_\chi\partial_\psi R_{\rm c}}{h_\psi \partial_\chi R_{\rm c}} \right )^2 \, \right ]^{1/2}_{\chi_{\rm c}}
 (\hat{r} \cdot \hat{\chi})_{\chi_{\rm c}}  \int_0^{\chi_{\rm c} (\psi)} d\chi h_\chi h_\varphi h_\psi \rho_\rx \varpi^2.
 \label{Itx}
\ee
where $\rx=\{\rp,\,\rn\}$ for protons and neutrons respectively.
On the other hand, the $\mbox{x}$-fluid's total moment of inertia in magnetic coordinates is given by
\be
I_\rx = \int_{r\leq R_{\rm c}} d\chi dS h_\chi (\hat{r} \cdot \hat{\chi})  \rho_\rx \varpi^2,
\ee
which, after carrying out the $\varphi$-integration and using the upper-lower hemisphere symmetry, becomes\footnote{It is worth noting that, in the limiting case
of cylindrical coordinates such that $ ( \psi,\chi,\varphi ) \to ( \varpi, z, \varphi ) $, we have $h_\psi = h_\chi =1$ and 
$\hat{r} \cdot \hat{\chi} = \cos\theta = z/r$, and (\ref{Ixmag}) reduces to (\ref{Ixcyl}). },
\be
I_\rx = 4\pi \int d\psi \int_0^{\chi_{\rm c} (\psi)}  d\chi h_\varphi h_\psi h_\chi (\hat{r} \cdot \hat{\chi})  
\left [ 1 + \left ( \frac{h_\chi\partial_\psi r}{h_\psi \partial_\chi r} \right )^2 \, \right ]^{1/2}  \rho_\rx \varpi^2. 
\label{Ixmag}
\ee
It follows from (\ref{Itx}) and (\ref{Ixmag}), together with the geometric identity (\ref{eqn:geomident}), that 
\be
	\tilde{I}_\rx = I_\rx.
\ee 
We therefore conclude that a general axisymmetric dipole poloidal 
field (with no closed lines) couples the \textit{entire} charged fluid core to the crust, thus enforcing corotation between these two components.  

This result does not support Easson's claim that poloidal fields (with open lines) do not always enforce crust-core corotation~\citep{easson79}. 
It is not difficult to find the reason behind this disagreement:  as we have already pointed out, the counter-example calculation provided by Easson 
makes use of the unphysical interior dipole field $f=1/r$.

%%%%%%%%%%%%%%%%%%%%%%%%%%%%%%%%%%%%%%%%%%%%%%%%%%%%%%%%%%
%%%%%%%%%%%%%%%%%%%%%%%%%%%%%%%%%%%%%%%%%%%%%%%%%%%%%%%%%%%

\section{Crust-core corotation: mixed poloidal-toroidal fields}
\label{sec:mixed}

The preceding purely poloidal background field model is an ideal testbed for understanding and solving the
magnetic crust-core corotation problem. However, based on arguments about dynamical stability, it is commonly believed that realistic 
neutron stars are likely to harbour a field of mixed poloidal-toroidal geometry. Therefore, it is highly desirable to extend the previous 
analysis to encompass  the case of a mixed magnetic field. 

\begin{figure}
\centerline{
\includegraphics[height=7cm,clip]{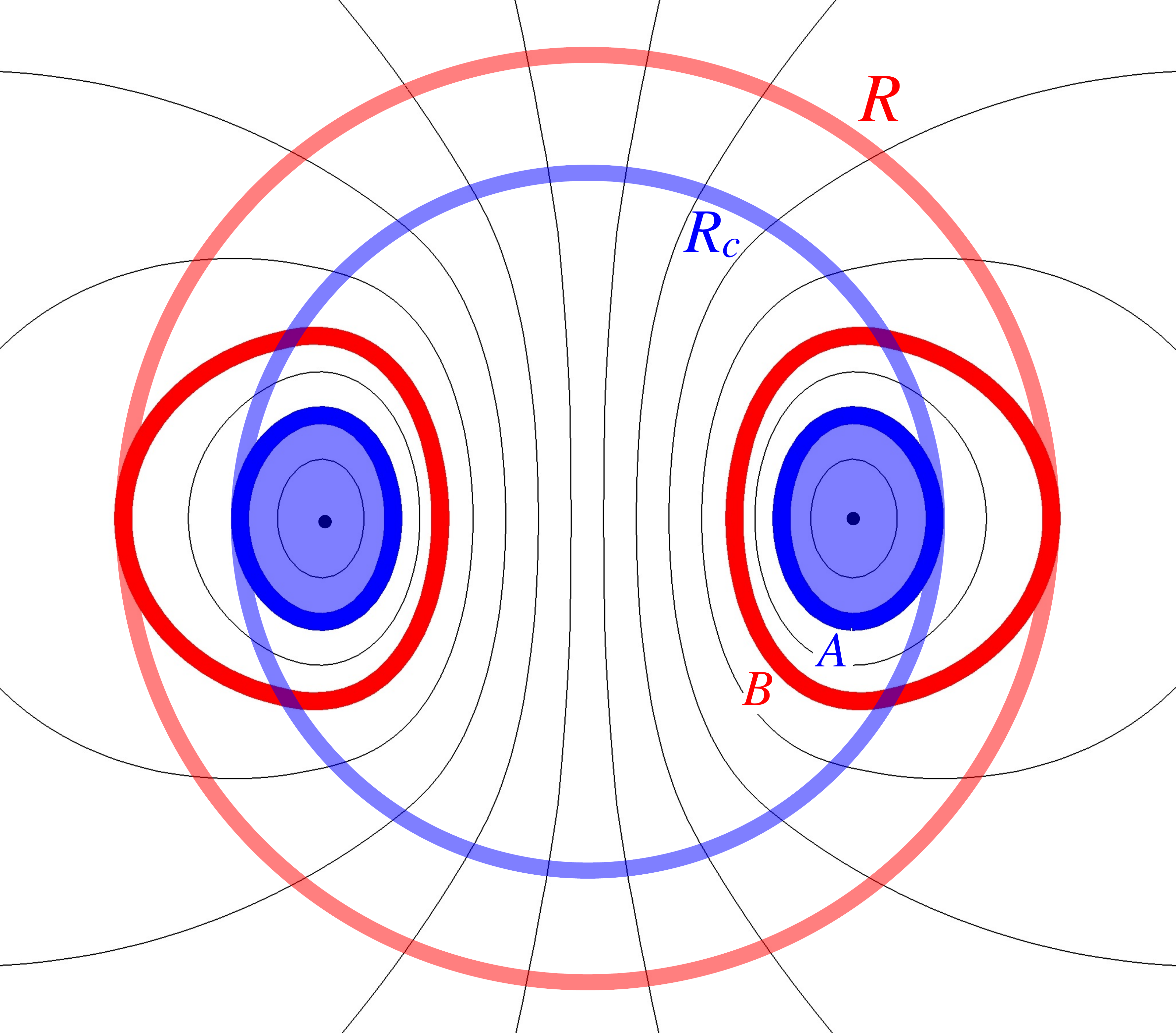}
\includegraphics[height=7cm,clip]{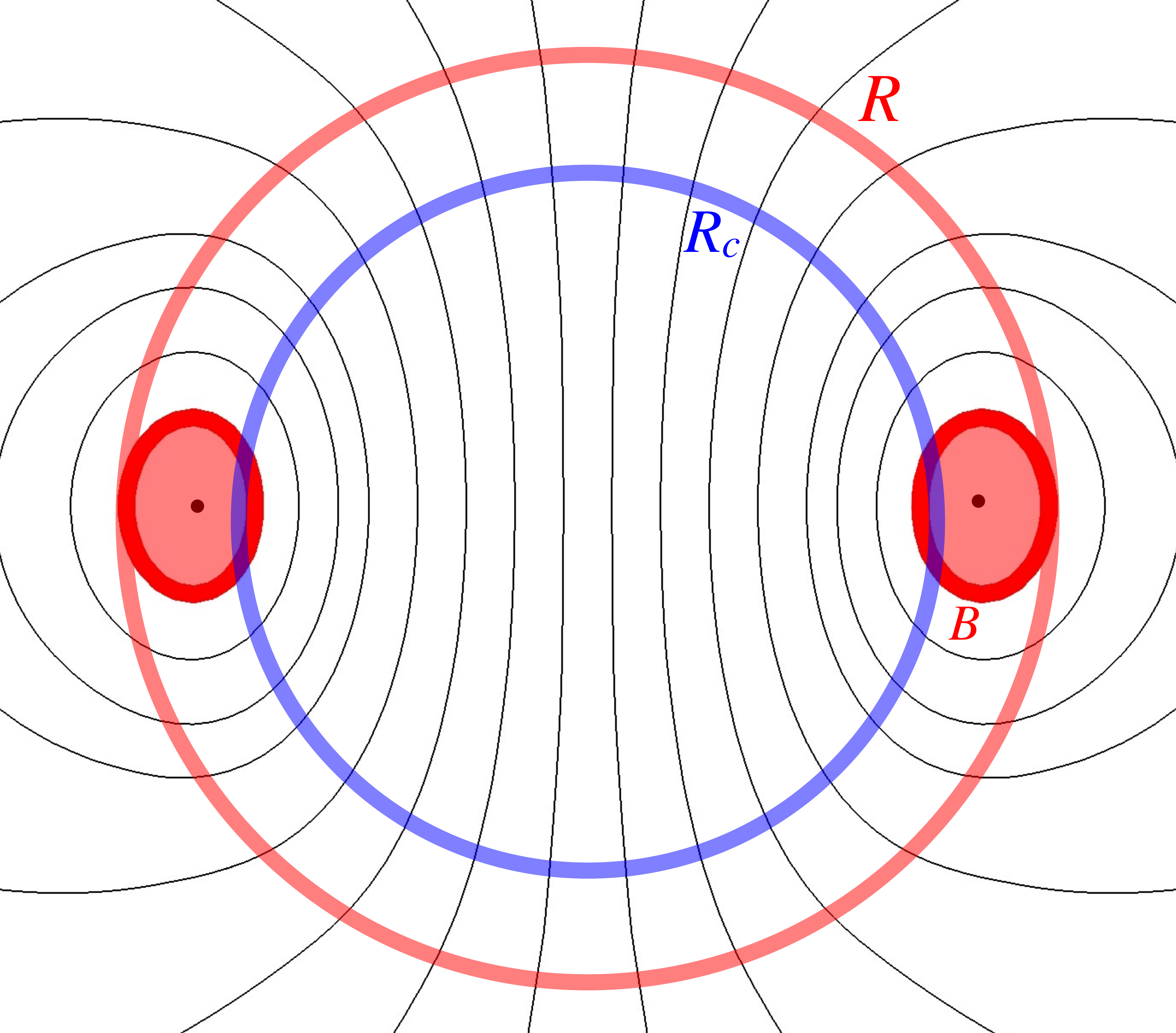}
}
\caption{Schematic configurations of mixed poloidal-toroidal magnetic fields.  The outer crust and crust-core interface are the red and blue circles labelled $R$ and $R_{\rm c}$ respectively.  Left panel: closed field lines in the core --- the thick blue field line, labelled A, is the outer-most poloidal field line that is closed in the core, and the red magnetic field line, labelled B, is the outer-most closed field line in the entire star.  For the twisted-torus model, toroidal field is confined to the region enclosed by B.  The black dots represent the neutral line (the point where the poloidal magnetic field component is zero).  Right panel: no closed field lines in the core -- the only closed field lines penetrate the crust-core interface, with B being the outer-most closed poloidal field line.  Again, the twisted-torus model has toroidal field confined to the region enclosed by B.}
\label{fig:Bfield}
\end{figure}

The assumed geometry of the field is displayed in Fig.~\ref{fig:Bfield} and makes contact with the twisted torus
configuration discussed earlier. More specifically, the poloidal field threads the entire star and has a neutral point (line) somewhere in the core. 
The toroidal field is confined inside the star, occupying the region around the neutral point, and is bounded by the last closed field line inside 
the star, i.e., line B in the figure. On the other hand, the line A marks the last closed field line in the core. 
As a result, the mixed poloidal-toroidal field lines that thread both the crust and core are the ones between the lines A and B.

%%%%%%%%%%
%%%%%%%%%%%

As we have already seen, the spindown quasi-equilibrium for a mixed background field, $\mathbf{B}_0$, is described by eqn.~(\ref{Ephi1}),
\be
\mathbf{B}_{\rP} \cdot \nabla (\varpi b^\varphi) + \mathbf{b}_\rP \cdot \nabla (\varpi B^\varphi_0) =  
4\pi  \varpi^2 \left ( \, - \rho_\rp \dot{\Omega}  + 2 \cB \rho_\rn \Omega_\rn \Omega_\rnp \, \right ).
\label{mix1}
\ee
Since $\mathbf{b}_\rP$ is axisymmetric, we can introduce a new stream function $\beta (r,\theta)$:
\be
\mathbf{b}_\rP = \nabla \beta \times \nabla \varphi.
\ee 
Expressed in magnetic coordinates, where $\beta=\beta(\psi,\chi)$, 
\be
\mathbf{b}_\rP = \frac{\partial_\psi \beta}{h_\varphi h_\psi} \hat{\chi} - \frac{\partial_\chi \beta}{h_\varphi h_\chi} \hat{\psi}.
\ee
We also introduce a stream function $\cI$ for the background toroidal field,
\be
B^\varphi_0 = \cI \nabla \varphi.
\ee
It is well known that for an axisymmetric magnetic equilibrium the two stream functions are related as $\cI = \cI (\psi)$ (e.g. \citet{GAL}).

In terms of these new functions, we can write (\ref{mix1}) as (recall that $B_\rP= 1/h_\psi h_\varphi$, see Appendix~\ref{sec:coordinates})
\be
\partial_\chi T - \cI^\prime \partial_\chi \beta = 4\pi \varpi^2  h_\varphi h_\chi h_\psi 
\left ( \, -  \rho_\rp \dot{\Omega}  +  2 \cB  \rho_\rn \Omega_\rn \Omega_\rnp  \, \right ).
\label{mix2}
\ee
It is straightforward to see that (\ref{mix2}) breaks down when integrated along a closed field line in the core. 
Along such a line $\cI^\prime$ is constant and therefore the integration leads to
\be
0= \oint d\chi  \varpi^2 h_\varphi h_\chi h_\psi \left ( \, -\rho_\rp \dot{\Omega}  + 2 \cB \rho_\rn \Omega_\rn \Omega_\rnp  \, \right ), 
\ee
which is a contradiction. This verifies the assertion discussed earlier in Section~\ref{sec:corotation} that closed field lines are 
incompatible with the assumption of crust-core corotation.

Next, we consider (\ref{mix2}) along an open field line. Its integration yields 
\be
M (\chi,\psi) \equiv T(\chi,\psi) - \cI^\prime \beta (\chi,\psi) 
=  -4\pi \dot{\Omega} \int_0^\chi d\chi^\prime  h_{\chi^\prime} h_\psi h_\varphi  \rho_\rp \varpi^2
+ 8\pi \cB \Omega_\rnp \Omega_\rn \int_0^\chi d\chi^\prime  h_{\chi^\prime} h_\psi h_\varphi \rho_\rp \varpi^2,
\ee
where we have used the assumed equatorial symmetry and anti-symmetry for $b^r$ and $b^\varphi$, respectively\footnote{A symmetric $b^r$ 
entails an antisymmetric function $\beta$; this becomes evident since in the limit $z \to 0$ we have $b^\psi \to b^r$ and $\partial_\chi \to \partial_z$.}.

For the crust-core torque we have (remember that this is the contribution of the upper hemisphere multiplied by two):
\be
N_{\rm cc} = \frac{1}{2\pi} \int_{R_{\rm c}} \frac{dS}{\varpi_{\rm c}} \left [\, \left  \{\, M + \partial_\psi (\cI \beta) \, \right  \} \frac{(\hat{r} \cdot \hat{\chi})}{h_\psi} 
- \cI \frac{\partial_\chi \beta}{h_\chi} ( \hat{r} \cdot \hat{\psi})  \, \right ].
\label{Nmix1}
\ee
In taking the surface integral, we need to make sure that the range of $\psi$ is limited to the open lines across the crust-core interface.  That is, 
$\psi$ should `run' between the value $\psi_z$ of the symmetry axis and $\psi_A$ of the line A, see Fig.~\ref{fig:Bfield}. 
At the same time,  the range for the $\psi$-integration of the terms that depend on the toroidal field $\cI$ is the subset  $ \psi_B < \psi <\psi_A$ 
of the previous range, with $\psi_B$ corresponding to the boundary line B of the toroidal field region. 

It is straightforward to obtain,
\be
N_{\rm cc} = -\bar{I}_\rp \dot{\Omega}  + 2\cB \Omega_\rn \Omega_\rnp \bar{I}_\rn  
+ \int_{\psi_B}^{\psi_A} d\psi 
\left [ 1 + \left ( \frac{h_\chi\partial_\psi R_{\rm c}}{h_\psi \partial_\chi R_{\rm c}} \right )^2 \, \right ]^{1/2}_{\chi_c}
\left \{\, \partial_\psi ( \cI  \beta )  (\hat{r} \cdot \hat{\chi}) 
- \cI \partial_\chi \beta \frac{h_\psi}{h_\chi}   (\hat{r} \cdot \hat{\psi})  \,\right \}_{\chi_c},
\label{Nmix2}
\ee
where we have defined
\be
\bar{I}_\rx = 4 \pi \int_{\psi_z}^{\psi_A} d\psi\,   \left [ 1 + \left ( \frac{h_\chi\partial_\psi R_{\rm c}}{h_\psi \partial_\chi R_{\rm c}} 
\right )^2 \, \right ]^{1/2}_{\chi_{\rm c}}
 (\hat{r} \cdot \hat{\chi})_{\chi_{\rm c}}  \int_0^{\chi_{\rm c} (\psi)} d\chi h_\chi h_\psi h_\varphi \rho_\rx \varpi^2.
 \label{Ixbar1}
\ee
This can be compared with the moment of inertia $\tilde{I}_\rx$ appearing in the earlier calculation of the poloidal field with
open lines. The difference in the integration limits implies that  $\bar{I}_\rx \leq  \tilde{I}_\rx$, with the equality holding when there are no
closed lines in the core or, equivalently, when the poloidal's field neutral point is pushed outside into the crust. 

So far the discussion in this section applies to the case of an arbitrary twisted torus-like field, but in order to proceed further we 
make the assumption of a dipole poloidal field $\cB_\rP$, i.e. 
\be
\psi = A f(\tilde{r}) \sin^2 \theta, \qquad \tilde{r} = \frac{r}{R},
\ee
where for convenience we have taken $f$ to be a dimensionless function. In Appendix~\ref{sec:geom}
we obtain the following results for the geometric factors appearing in (\ref{Nmix2}) and (\ref{Ixbar1}): 
\be
\left [ 1 + \left ( \frac{h_\chi\partial_\psi R_{\rm c}}{h_\psi \partial_\chi R_{\rm c}} \right )^2 \, \right ]^{1/2}  
(\hat{r} \cdot \hat{\chi}) = 1, \qquad 
\frac{h_\psi}{h_\chi} \left [ 1 + \left ( \frac{h_\chi\partial_\psi R_{\rm c}}{h_\psi \partial_\chi R_{\rm c}} \right )^2 \, \right ]^{1/2}  (\hat{r} \cdot \hat{\psi}) 
= \pm \frac{G}{2A f \cos\theta},
\label{geom1}
\ee
where the explicit form of the function $G (r) $ is given in Appendix~\ref{sec:coordinates}. The sign arbitrariness of the second equation 
comes from the fact that the inner product $\hat{r} \cdot \hat{\psi}$ can switch sign along a given field line (open or closed). 

We then have from (\ref{Nmix2}) and (\ref{Ixbar1}),
\be
\bar{I}_\rx = 4 \pi \int_{\psi_z}^{\psi_A} d\psi\, \int_0^{\chi_{\rm c} (\psi)} d\chi h_\chi  h_\psi h_\varphi \rho_\rx \varpi^2  \leq I_\rx,
\label{Ixbar2}
\ee
and 
\be
N_{\rm cc} = -\bar{I}_\rp \dot{\Omega}  + 2\cB \Omega_\rn \Omega_\rnp \bar{I}_\rn  \\
+  \cI (\psi_A) \beta (\chi_{\rm c},\psi_A) 
 \mp \frac{G_{\rm c}}{2A f_{\rm c}}  \int_{\psi_B}^{\psi_A} d\psi  \cI \frac{(\partial_\chi \beta)_{\chi_c}}{\cos\theta},
\label{Nmix3}
\ee
where  $G_{\rm c}, f_{\rm c}$ are the values of $G,f$ at $r=R_{\rm c}$. We have also used the fact that $ \cI (\psi_B)=0$
for the assumed magnetic field geometry. With the help of (see Appendix~\ref{sec:coordinates})
\be
\chi = G \cos\theta \quad \to \quad \chi_{\rm c} (\psi) = G_{\rm c} \left ( \, 1-\frac{\psi}{A f_{\rm c}} \,  \right )^{1/2},
\ee
we can obtain the equivalent expression,
\be
N_{\rm cc} = -\bar{I}_\rp \dot{\Omega}  + 2\cB \Omega_\rn \Omega_\rnp \bar{I}_\rn  
+ \cI (\psi_A) \beta (\chi_{\rm c},\psi_A)  
\mp \frac{G_{\rm c}}{2A f_{\rm c}} \int_{\psi_B}^{\psi_A} d\psi   \left [ \, 1-\frac{\psi}{A f_{\rm c}} \,  \right ]^{-1/2}   
\cI (\partial_\chi \beta)_{\chi_{\rm c}}.
\label{Nmix4}
\ee
In order to proceed further, an independent solution of the form $\beta = \alpha_1 \dot{\Omega} + \alpha_2 \Omega_\rnp \Omega_\rn$
must be provided; such a result can only be obtained by solving the remaining poloidal components of the Euler equation~(\ref{eulerp}).
These equations can be found in Appendix~\ref{sec:Epol}, however, solving them analytically does not appear to be possible.

This difficulty did not crop up in our earlier poloidal field analysis because in that case the $\varphi$-Euler equation,
the torque $N_{\rm cc}$ and $b^\varphi$ itself decouple from $\mathbf{b}_\rP$ and the poloidal Euler equations. In practise, this means
that a purely poloidal $\mathbf{B}_0$ field is compatible with the toroidal spindown `mode' $(b^\varphi,\mathbf{b}_\rP,\delta \Psi_\rp)= (b^\varphi,0,0)$.  

A purely toroidal spindown mode is no longer admissible when $B_0^\varphi \neq 0$ but, nevertheless, it is still possible to find a 
mixed toroidal-poloidal mode that would allow a complete calculation of the crust-core torque $N_{\rm cc}$.  The desired $\mathbf{b}$ is simply 
the superposition\footnote{In terms of a multipole expansion in Legendre polynomials $P_\ell (\cos\theta)$ this superposition would 
correspond to an $\ell=2,4,6, ...$ sum for $b^\varphi$ and an $\ell=1,3,5,...$ sum for $\mathbf{b}_\rP$.}:
\be
\mathbf{b} = b^r_{\rm a} \hat{r} + b^\theta_{\rm s} \hat{\theta} +  b^\varphi_{\rm a} \hat{\varphi},
\label{sup}
\ee
where `a' and `s' stand, respectively, for antisymmetric and symmetric.

As a result of the imposed symmetries we have that (i) the $\sim b^r B_0^\varphi$ term in the crust-core torque (\ref{N1}) integrates to
zero and (ii) the second left-hand-side term in (\ref{Ephi1}) is anti-symmetric (as opposed to the other symmetric terms) 
and therefore should be zero, i.e. $\mathbf{b}_\rP \cdot \nabla (\varpi B_0^\varphi) =0 \to \cI^\prime \partial_\chi \beta =0 $. 
In other words, this spindown mode is characterized by a poloidal perturbation that points along the background field lines,  
$\mathbf{b}_\rP = b^\chi \hat{\chi}$. At the same time the perturbation (\ref{sup}) also leads to a pair of non-trivial poloidal Euler 
equations for $\beta$ and $\Psi_\rp$. 

Adopting (\ref{sup}) as a spindown `mode' implies that we can remove all $\beta$-dependence from the general result (\ref{Nmix3}) (or (\ref{Nmix4})),
leaving 
\be
N_{\rm cc} = -\bar{I}_\rp \dot{\Omega}  + 2\cB \Omega_\rn \Omega_\rnp \bar{I}_\rn,
\label{Nmix5}
\ee
as the main result for the total crust-core torque for the case of a mixed magnetic field. According to this, the magnetic field
couples the crust with exactly that portion of the core that is threaded by the open magnetic field lines. Assuming that poloidal field has
a neutral point somewhere in the core (as in the left panel of Fig.~\ref{fig:Bfield}), the toroidal-shaped region
between the field lines A and B (i.e. the shaded region in Fig.~\ref{fig:Bfield}) remains magnetically decoupled and of course makes no contribution 
to the torque. The same property can be seen to hold true for the more general case represented by eqn.~(\ref{Nmix4}). 
The astrophysical implications of this remarkable result are discussed in the following section.  

Finally, it is worth remarking on a fundamental difference between the above argument regarding the crust-core torque and Easson's no-crust-core corotation 
argument. We remind the reader that \citet{easson79} showed that a crust-core corotation equilibrium \textit{cannot} exist  along closed poloidal field lines in 
the core, as the governing equations lead necessarily to a mathematical inconsistency when corotation is assumed (see Section \ref{sec:corotation}).  
Our analysis, facilitated by the use of magnetic coordinates, has shown that the closed field lines of poloidal/mixed magnetic fields do not couple 
at all with the crust and as a result the crust cannot  exert any torque on the region occupied by the closed lines. These two anti-corotation arguments are clearly complementary; our analysis explains that closed field lines can violate the corotation assumption as they are decoupled from the crust.  
This subtle issue may be purely academic; as we discuss below, we expect the spin-lag generated between the open- and closed-field-line regions to render the latter region dynamically unstable, implying crust-core corotation can only be established in the absence of closed field lines in the core.

%%%%%%%%%%%%%%%%%%%%%%%%%%%%%%%%%%%%%%%%%%%%%%%%
%%%%%%%%%%%%%%%%%%%%%%%%%%%%%%%%%%%%%%%%%%%%%%%%

\section{Failure of long-term crust-core corotation: implications}
\label{sec:implications}

\subsection{Incomplete crust-core coupling: how likely is it?}

The results obtained in the preceding sections strongly suggest that the extent to which the liquid core can follow the secular
spindown of the solid crust is largely decided by the geometry of the magnetic field \textit{in the core}, in agreement with the original conjecture 
put forward by \citet{easson79}. A key question raised in that early work concerned the `size' of the class of magnetic fields  that
are unable to enforce a full crust-core coupling, and hence allow for the build up of a persistent spin-lag between part of the core and the rest
of the star.

This paper provides a definitive answer to that crucial question.  As we have seen, an arbitrary dipole poloidal field with no closed
field lines in the core does enforce complete crust-core corotation, in agreement with  `conventional wisdom'. However, the presence of 
purely poloidal magnetic fields in neutron stars enjoys very little theoretical support since these fields are inherently unstable (e.g. \citet{wright73, FR77,
braithwaite07,lasky11}) (the same is true for purely toroidal fields \citep{tayler73,braithwaite06c}). Instead, hybrid poloidal-toroidal  `twisted torus' fields 
are the ones commonly attributed to realistic neutron stars and the ones that have attracted most theoretical work in the recent years (e.g.
\citet{braithwaite06a,colaiuda08,ciolfi09,mastrano11,GAL,lander12,CR13,fujisawa14}). Such fields are exemplified by the left-hand panel of Fig.~\ref{fig:Bfield}: 
they typically have closed field lines in the core (although the neutral line may be located in the crust for sufficiently strong toroidal components \citep{ciolfi09}) and, as we have shown in Section~\ref{sec:mixed},  the part of the core occupied by these lines remains magnetically 
decoupled from the crust (and the rest of the core). 

This result implies that the aforementioned class of magnetic fields is `large', in the sense that such fields should be considered likely to occur in real systems.  
These include the most commonly used twisted torus models, but not necessarily fields for which the magnetic field is misaligned with the rotation axis (see section \ref{sec:conclusions}). In what follows we discuss the spindown evolution of a neutron star endowed with a twisted 
torus-like magnetic field and speculate on the implications of our results.

%%%%%%%%%%%%%%%%%%%%%%%%%%%%%%%%%%%%%%%%%%%%%%%%%%

\subsection{Implications for the magnetic field evolution}

To what extent the crust and the core can be interlocked in long-term corotation is most relevant 
soon after a neutron star's birth. Assuming a standard cooling scenario, the crust is expected to form approximately a day
after the neutron star is formed, when the temperature has dropped to $ T \sim \mbox{(few)} \times 10^9\,\mbox{K}$ \citep{STbook}.
During the same period, the star retains most of its initial kinetic energy (this is true even for newly-born magnetars unless their surface
dipole field exceeds approximately $10^{14}-10^{15}\,\mbox{G}$) and therefore most of the subsequent electromagnetic spindown 
will take place when the star has a well defined solid crust and a liquid core. The onset of neutron superfluidity in the core is likely to ensue much 
later ($\sim 1\,\mbox{yr} - 100\,\mbox{yr}$), when the star has cooled below $T \sim 10^9\,\mbox{K} -\mbox{(few)} \times10^8\,\mbox{K}$, respectively \citep{shterninetal11,pageetal11}.  On the other hand, the formation of the superconducting proton condensate is expected to take place much earlier, 
at $T > 10^9\,\mbox{K}$~\citep{STbook}.
  
Once the crust is well formed and has begun to be torqued by the anchored magnetic field, two mechanisms come into play in an attempt to enforce crust-core 
corotation: the magnetic field and hydrodynamical viscous Ekman pumping. The latter mechanism has been extensively discussed in the neutron star literature, 
most notably in \citet{easson79,abney,melatos12}, see also \citet{Greenspan}. As recently emphasized in \citet{melatos12}, the combined effect of stratification and compressibility in neutron star matter severely inhibits the efficiency of the Ekman mechanism, resulting in corotation timescales much longer than initially anticipated \citep{abney}. For typical neutron stars parameters, a viscous coupling timescale as long as $\sim 10^3\,\mbox{yr}$ is expected \citep{melatos12}. The magnetic coupling mechanism is much faster than this (see also our discussion in Section~\ref{sec:intro}); we can therefore frame our discussion without having to consider viscosity, consistent with the model presented in this paper.

A day-old neutron star is likely to start its secular spindown evolution being in stable hydromagnetic equilibrium with a magnetic field geometry that resembles 
the ones shown in Fig.~\ref{fig:Bfield}. Consider first the case of the right panel of Fig.~\ref{fig:Bfield}. In that scenario, the poloidal neutral point is located in the crust,
as is the bulk of the toroidal field. The entire core is threaded by open (poloidal and toroidal) field lines and, as we have shown in this paper, the magnetic 
coupling can enforce corotation between the crust and the total amount of charged  fluids in the core. Our analysis and results apply equally well to a system with 
normal or superfluid neutrons, in the former case  the implication being that the magnetic field directly couples all the fluids to the crust, while in the latter case 
vortex mutual friction is very efficient in enforcing corotation between the superfluid and the protons. 

More interesting (and perhaps more probable) is the case of neutron stars `starting' with the magnetic field equilibrium 
of the left panel of Fig.~\ref{fig:Bfield}, where the poloidal neutral point and the bulk of the toroidal field are located in the outer core. 
In such a case, the toroidal-shaped region bounded by the closed line A (depicted as a shaded region in Fig.~\ref{fig:Bfield}) remains  
magnetically decoupled to the crust and, as the rest of the star spins down, a spin lag will begin to build up, with the toroidal region 
retaining its initial `fossil' spin frequency.   

It is unlikely that this state of persistent crust-core spin lag can remain dynamically stable for long. The velocity jump in the proton fluid across 
boundary A will give rise to an increasingly strong azimuthal current sheet $\Delta J \hat{\varphi}$ and an associated poloidal Lorentz force 
$\Delta J \hat{\varphi} \times \mathbf{B}_\rP /c$. We would expect $|\Delta J / J| \sim \Delta \Omega/\Omega $ where 
$\Delta \Omega (t) \sim \dot{\Omega} t$ is the spin lag. This means that the induced Lorentz force can seriously disturb the background equilibrium after 
a time $t \sim t_{\rm sd} = \Omega/|\dot{\Omega}|$. 

The physical conditions along the same boundary field line A are also suitable for the onset of the classical Kelvin-Helmholtz instability. 
This should take place once the spin lag exceeds the threshold  $\Delta \Omega_{\rm KH} \sim v_\rA /R$, where $v_\rA$ is the average Alfv\'en speed of 
the local toroidal field $B^\varphi_0$ (see \citet{chandra}, Chapter XI, p. 507).
The corresponding timescale for the onset of the instability, $t_{\rm KH}$, can be estimated to be (here we use canonical values $M=1.4\, M_\odot$
and $R=10\,\mbox{km}$ for the stellar mass and radius):
\be
\frac{t_{\rm KH}}{t_{\rm sd}}  \sim \frac{v_\rA}{\Omega R}  \approx 5\times 10^{-2} \left ( \frac{B^\varphi_0}{10^{15}\,\mbox{G}} \right ) 
\left ( \frac{P}{10\,\mbox{ms}} \right ),
\label{tKH}
\ee
suggesting that it can set in much sooner than the lapse of one spindown timescale. The inclusion of proton superconductivity  
can significantly raise the above ratio, $v_\rA / \Omega R \approx 0.2\, (B^\varphi_0/10^{15}\,\mbox{G})^{1/2} (P/10\,\mbox{ms}) $,  
but this is still well below unity provided $P \lesssim 10\,\mbox{ms}$, assuming a magnetar-strength field $B^\varphi_0$.

We thus have good reason to expect that the initial persistent crust-core spin lag will lead to some kind of magnetohydrodynamic instability
after a time (assuming a standard spindown torque, see e.g.~\citet{STbook})
\be
t \lesssim t_{\rm sd} \approx 4.7\, \left ( \frac{B_d}{10^{15}\,\mbox{G}} \right )^{-2} \left ( \frac{P}{10\,\mbox{ms}} \right)^2\, \mbox{d},
\ee
where $B_d$ is the dipole surface field. 
Among all systems, newly born magnetars appear to be the most favoured ones with respect to the onset of the instability,
typically having  $t_{\rm sd}  \sim 1  \mbox{d} - 1\,\mbox{yr}$. On the other hand, however, too strongly magnetized and/or too rapidly rotating 
magnetars will spin down before the formation of their crusts, rendering this discussion moot.

It is beyond the scope of this paper to provide a detailed analysis of the fate of the unstable, super-rotating core region.  Nevertheless, we can speculate.
Accepting that an initial magnetic equilibrium such as that in the left panel of Fig.~\ref{fig:Bfield}
is quickly rendered dynamically unstable by the super-rotation of the closed-field-lines region, the magnetic field will try to rearrange itself and find a new stable
equilibrium. As long as the neutral line remains in the core, this does not seem viable; any new configuration will still contain closed field line regions which will 
again lead to a crust-core spin-lag.

Thus, a stable equilibrium is likely to be reached only by the eviction of the neutral line and the attached toroidal field from the core and  
into the crust.  We therefore suggest that the failure of the magnetic field to enforce complete crust-core corotation drives a magnetic field evolution
in very young neutron stars  akin to moving from \textit{the left to the right panel} of Fig.~\ref{fig:Bfield}.

This hypothetical evolutionary scenario sees the crust as the natural \textit{depository} of toroidal magnetic field and energy initially located in the core.
As pointed out earlier, this kind of rotation-driven magnetic field evolution could be especially favoured in magnetars because these magnetized objects 
experience a rapid spindown soon after the formation of the crust. In other words, and according to the above suggestion, magnetars may
have their strong interior toroidal fields pushed into their crusts in $\sim 1\,\mbox{d}-1\, \rm{yr}$ after their birth. This prediction sits well with our theoretical 
understanding of magnetar thermal properties and seismic activity. A strong, evolving toroidal field in the crust can easily provide the energy required to 
heat magnetar surfaces to the observed levels \citep{pons07,kaminkeretal06,HGA12}. The same process could be responsible for fracturing the crust 
and powering high energy events like magnetar flares \citep{TD01}. 

At this stage it is more difficult to speculate on the repercussions of the proposed magnetic field evolution on canonical systems like radio pulsars,
or the less magnetized central compact objects (CCOs).  The smoother electromagnetic spindown of these objects suggests a much longer spin 
lag longevity, as $t_{\rm sd} \sim 10^3\,\mbox{yr}-10^4\,\mbox{yr}$. It should be recalled that the viscous coupling timescale can be of comparable 
length \citep{melatos12} and therefore corotation could be achieved regardless of magnetic coupling. 
It is more likely, however, that the Kelvin-Helmholtz instability will drive magnetic field evolution much earlier than that, especially if the internal toroidal field 
is comparable to the surface field, i.e. $B^\varphi_0 \sim 10^{12}\,\mbox{G}$; see equation (\ref{tKH}). In this latter case, and accounting for superconductivity, 
the instability would set in after $\sim 1\,\mbox{yr}-100\,\mbox{yr}$.

Although the previous scenario is attractive for all the reasons discussed, it is worth reiterating that it is based mainly in conjecture.  
However, if for some reason it is not realised, it is possible that a persistent `quasi-steady' turbulent state could be established, driven 
by shear flows at the interface between the two regions.  Turbulent flows in the interiors of neutron stars have many potentially interesting 
observational consequences \citep[for example see][and references therein]{mastrano05,melatos12}. 
Establishing whether the closed-field-line region is evicted from the core or not is a difficult task that we anticipate will require high-resolution 
numerical simulations to resolve.

Our final remark concerns the role of neutron superfluidity in the evolution of the super-rotating core region. This could be relevant if the persistent spin lag 
survives until the bulk of the stellar core has become superfluid ($\sim 1\,\mbox{yr} - 100\,\mbox{yr}$). Given the previous (very approximate!) instability 
timescales, this scenario is more likely to be realized in young pulsars, although the low magnetic field/spin frequency end of the magnetar distribution cannot
be excluded. The neutron vortex lattice which enables the superfluid to rotate will generally thread both the core region coupled to the crust and the decoupled super-rotating region. The mutual friction force will efficiently couple the neutrons to the protons in both regions, giving rise to a differentially rotating superfluid. In order for 
this to be possible, the vortices that thread both regions must adapt their spatial distribution and, at the same time, bend in the vicinity of the boundary field line A 
(see Fig.~\ref{fig:Bfield}). However, the self-tension of the vortices will prevent this from happening indefinitely. In fact, it might be possible that corotation is established
in the entire core under the combined action of the magnetic field and the superfluid vortices. This is an attractive scenario but is one that, as we have pointed out,
requires an initially long-lived persistent crust-core spin lag.

%%%%%%%%%%%%%%%%%%%%%%%%%%%%%%%%%%%%%%%%%%%%%%%%%%%%%%%%%%%%%%%%%%
%%%%%%%%%%%%%%%%%%%%%%%%%%%%%%%%%%%%%%%%%%%%%%%%%%%%%%%%%%%%%%%%%%

\section{Concluding discussion}
\label{sec:conclusions}

In this paper, we derive conditions for a magnetic field threading the interior of a neutron star to enforce long-term corotation between the slowly-spinning down crust of the star and the superfluid core.  We show that a magnetic field with open field lines allows for complete corotation between the two components.  However, somewhat counter-intuitively, the presence of any closed field-line region in the core of the star causes that region to be magnetically decoupled from the rest of the star.  As the crust and the open-field-line region of the core spin down in unison, a velocity lag will build up between the closed-field-line region and the rest of the star.  

Example magnetic field topologies that will generate such persistent spin lags include the popular `twisted torus' configurations, in which toroidal field threads the closed-field-line region of the internal poloidal field (see the left hand panel of Fig.~\ref{fig:Bfield}).  Our results indicate that, soon after the formation of the crust, a young neutron star with, for example, a twisted torus magnetic field will develop a persistent spin lag in the region enclosed by boundary A in the left panel of Fig.~\ref{fig:Bfield}.  The long-term fate of this super-rotating core region is unclear.  The velocity difference between the two regions gives rise to a current, and a corresponding Lorentz force that needs to be considered when calculating new magnetohydrostatic equilibria.  However, we expect any such equilibria to be unstable to e.g., the classical Kelvin-Helmholtz instability at the interface between the two regions (boundary A in the left-hand panel of Fig.~\ref{fig:Bfield}).  In Section \ref{sec:implications}, we speculate that any long-term evolution would involve the eviction of the toroidal component of the magnetic field into the crust, thereby allowing the entire core to corotate with the crust.

Fully addressing the question of the long-term fate of the super-rotating core region is a challenging task that requires sophisticated numerical simulations that, we believe, are beyond the capabilities of current techniques.  Some difficulties include: simultaneous resolution of the small-scale structure induced by the instabilities at the shear layer and the global magnetic field topology;  and the range of timescales in the system, from the secular spin-down and Kelvin-Helmholtz timescales ($\sim$ days -- years) down to the Alfv\'en timescale governing the dynamical adjustments of the magnetic field.  We therefore advocate a multipronged approach, include calculations of quasi-static, magnetohydrostatic evolutions to understand the influence of the poloidal Lorentz force on the toroidal-field-line region, as well as full magnetohydrodynamic simulations to understand the relevant instabilities at the shear layer.

Although our analysis makes no particular assumption about the properties of neutron star matter (equation of state, degree of stratification etc.), 
it does involve two main simplifications with respect to the magnetic field. The first one is the assumption of a dipolar structure for the poloidal field
which facilitated the full use of the magnetic coordinates formalism. Of course, these coordinates remain a well-defined notion and can be employed 
for the description of a general axisymmetric field, but this may come at the price of losing a large part of the analytic elegance of the dipole case. 
Although we have not pursued this issue further, it is unlikely that any of our conclusions will change for a magnetic field of more general, axisymmetric structure.
This may \textit{not} be true for non-axisymmetric fields where, for example, the poloidal field is misaligned with the stellar rotation axis \citep{lasky13}, 
however we suspect any field topology that admits closed field lines in the core will exhibit the same phenomenology. 
The second simplification is the use of the `classical' Lorentz magnetic force in the Euler equation. Strictly speaking this is `wrong' given that
the bulk of the proton fluid is expected to condense to a superconducting state around the same time as the formation of the crust. It is well known that
the magnetohydrodynamics of superconducting neutron stars can be quite different to that of their ordinary matter counterparts \citep{mendell98,supercon}.
One of the key differences\footnote{Another difference is the ``boosted'' superconducting Alfv\'en speed 
$v_\rA^2 = H_c B/4\pi \rho_\rp$; this expression has been used in some of our estimates in this paper.} is the form of the magnetic force, which in a 
superconducting system originates from the smooth-averaged self-tension of the quantized protonic fluxtubes. Future work should address this
approximation of our model by using the full machinery of superconducting magnetohydrodynamics.

%%%%%%%%%%%%%%%%%%%%%%%%%%%%%%%%%%%%%%%%%%%%%%%%%%%%

\section*{Acknowledgements}

We are grateful to Andrew Melatos for fruitful discussions and comments on the manuscript.
KG is supported by the Ram\'{o}n y Cajal Programme of the Spanish Ministerio de Ciencia e Innovaci\'{o}n and by the 
German Science Foundation (DFG) via SFB/TR7.  
PDL is supported by Australian Research Council Discovery Projects DP110103347 and DP1410102578.

%%%%%%%%%%%%%%%%%%%%%%%%%%%%%%%%%%%%%%%%%%%%%%%%%%%

\appendix

\section{Crust-core coupling for a non-uniform dipole field.}
\label{sec:fr4}

Here we consider the non-uniform dipole field $f(r) = r^4$ combined with the assumption of uniform density.
Then, eqn.~(\ref{pde1}) can be solved with separation of variables, leading to the solution
\be
b^\varphi (\theta)= \frac{\pi }{A\sqrt{\sin\theta}} \left ( \, - \rho_\rp \dot{\Omega}  + 2 \cB \rho_\rn \Omega_\rn \Omega_\rnp \, \right )
\int_\theta^{\pi/2}  d\theta^\prime \sqrt{\sin\theta^\prime}.
\ee
Using this result and $B^r_0 = 2A r^2 \cos\theta$ in the formula for the torque, $N_{\rm cc}$, we have after some rearrangement:
\be
N_{\rm cc}  = -\tilde{I}_\rp \dot{\Omega} + 2 \cB \Omega_\rn \Omega_\rnp \tilde{I}_\rn,
\ee
with the effective moments of inertia
\be
\tilde{I}_\rx = 2\pi R^5_{\rm c}  \rho_\rx \int_0^{\pi/2} d\theta \sin^2\theta \frac{\cos\theta}{\sqrt{\sin\theta}} 
 \int_\theta^{\pi/2}  d\theta^\prime \sqrt{\sin\theta^\prime}.
\ee
On the other hand, the total moment of inertia for each fluid is
\be
I_\rx = \int dV \rho_\rx r^2 \sin^2\theta = 4\pi \int_0^{R_{\rm c}} \int_0^{\pi/2} dr d\theta r^4 \sin^3 \theta \rho_\rx .
\ee
For uniform density this simplifies to
\be
I_\rx =  \frac{8\pi}{15} R_{\rm c}^5 \rho_\rx.
\ee
Our final task is to check the value of the ratio $\tilde{I}_\rx /I_\rx$. We have,
\be
\frac{\tilde{I}_\rx}{I_\rx} = \frac{15}{4} \int_0^{\pi/2} d\theta \sin^{3/2}\theta \cos\theta
 \int_\theta^{\pi/2}  d\theta^\prime \sqrt{\sin\theta^\prime}.
\ee
Direct numerical integration reveals that to numerical precision $\tilde{I}_\rx /I_\rx =1$ which implies a fully corotating core (protons and
electrons) with the crust.  

%%%%%%%%%%%%%%%%%%%%%%%%%%%%%%%%%%%%%%%%%%%%%%%%%%
%%%%%%%%%%%%%%%%%%%%%%%%%%%%%%%%%%%%%%%%%%%%%%%%%%

\section{Magnetic coordinates}
\label{sec:coordinates}

For a given axisymmetric poloidal field, $\mathbf{B}_\rP = \nabla \psi \times \nabla \varphi$, we can set up a system of magnetic 
coordinates, $\{ \psi,\chi,\varphi \}$, where $\chi$ measures distance along a field line (without loss of generality we can 
set $\chi=0$ at the magnetic equator).

The resulting coordinate system is orthogonal but curvilinear with line element,
\be
d\mathbf{r} \cdot d\mathbf{r}  = h_\psi^2 \,d\psi^2 + h_\chi^2 \,d\chi^2 + h_\varphi^2 \,d\varphi^2,
\ee
where $h_\psi, h_\chi, h_\varphi$ are all functions of $\psi$ and $\chi$, and it is easy to see that $h_\varphi = \varpi$.
By construction, $\mathbf{B}_\rP$ is tangent to $\psi$ surfaces, i.e., 
\be
\mathbf{B}_\rP = B_0 (\psi,\chi) \hat{\chi}.
\ee
The grad operator takes the form (where as always a `hat' denotes a unit vector)
\be
\nabla = \frac{\hat{\psi}}{h_\psi} \partial_\psi + \frac{\hat{\chi}}{h_\chi} \partial_\chi + 
\frac{\hat{\varphi}}{h_\varphi} \partial_\varphi.
\ee
We then find,
\be
\mathbf{B}_\rP = \nabla \psi \times \nabla \varphi = \frac{1}{h_\varphi h_\psi} \hat{\chi}. 
\label{Bpol_mc}
\ee

For the remainder of the analysis we assume a \textit{dipole} field, implying the stream function can be expressed in the general form
\be
\psi (r,\theta) = A f(\tilde{r}) \sin^2\theta, 
\label{stream3}
\ee
where $\tilde{r}=r/R$.  As we have seen, the components of this field in spherical coordinates are
\be
B_0^r = 2A \frac{f}{r^2} \cos\theta, \qquad B_0^\theta = -A\frac{f^\prime}{Rr} \sin\theta,
\ee
where $f^\prime = df(x)/dx$. 

From this we can obtain the metric component $h_\psi $:
\be
h_\psi  (r,\theta)  
= \frac{r}{A\sin\theta} \left [ \,   4 f^2\cos^2\theta + ( \tilde{r} f^\prime)^2 \sin^2\theta \, \right ]^{-1/2}.
\label{hpsi1}
\ee
Expressing the line element for the two-sphere in terms of $h_\psi$ and $h_\chi$, we can obtain three equations
\bear
&& \left ( h_\chi \frac{\partial \chi}{\partial r} \right )^2 = 1 - (\tilde{r} f^\prime)^2\sin^2\theta 
\left [ \,  4 f^2\cos^2\theta +  (\tilde{r} f^\prime)^2\sin^2\theta \, \right ]^{-1},
\label{eq1}
\\
\nonumber \\
&& \left ( h_\chi \frac{\partial \chi}{\partial \theta} \right )^2 = r^2 \left \{ 1 - 4f^2 \cos^2\theta 
\left [ \,  4 f^2 \cos^2\theta +  (\tilde{r} f^\prime)^2 \sin^2\theta \, \right ]^{-1} \right \},
\label{eq2}
\\
\nonumber \\
&&  h_\chi^2 \frac{\partial \chi}{\partial r} \frac{\partial \chi}{\partial \theta} =  - 2\sin\theta\cos\theta r\tilde{r} ff^\prime
\left [ \,  4 f^2\cos^2\theta + (\tilde{r} f^\prime)^2\sin^2\theta \, \right ]^{-1}.
\eear

It is straightforward to see that the bottom equation is trivially satisfied by the first two.  Simplifying the first two equations and taking square roots,
\bear
&& \frac{\partial \chi}{\partial r} = \pm 2f \frac{\cos\theta}{h_\chi} 
\left [ \,  4 f^2\cos^2\theta +  (\tilde{r} f^\prime)^2\sin^2\theta \, \right ]^{-1/2},
\label{dxdr1}
\\
\nonumber \\
&& \frac{\partial \chi}{\partial \theta} = \pm r \tilde{r} f^\prime \frac{\sin\theta}{h_\chi} 
\left [ \,  4 f^2 \cos^2\theta + (\tilde{r} f^\prime)^2 \sin^2\theta \, \right ]^{-1/2}.
\label{dxdth1}
\eear
Somewhat trivially, $\partial_{r}\chi>0$ for all $\chi(r,\theta)$ for a dipole poloidal field, and we therefore consider only the `$+$' sign in equation (\ref{dxdr1}).  
On the other hand, $\partial_\theta\chi$ can be positive and negative along any given field line.  For example, consider field line $A$ in the left hand panel of 
Fig.~\ref{fig:Bfield}.  As one moves from the innermost equatorial point on $A$ in the clockwise direction, $\partial_\theta\chi>0$ until the apex of the field line 
is reached, at which point $\partial_\theta\chi$ changes sign, and becomes negative as one moves further in the clockwise direction. 
We accordingly introduce $\epsilon=\pm1,\,0$ into equation (\ref{dxdth1}).

Motivated by solutions in the case when $f(\tilde{r})=\tilde r^a$ (see below), we search for a coordinate transformation of the form
\be
\chi(r,\theta) = G(r) \cos\theta.
\label{xansatz}
\ee
Inserting this ansatz into equations (\ref{dxdr1}) and (\ref{dxdth1}) gives respectively
\bear
&& \frac{dG}{dr} = \frac{2f}{h_\chi} \left [ \,  4 f^2 \cos^2\theta + (\tilde{r} f^\prime)^2 \sin^2\theta \, \right ]^{-1/2},
\label{dxdr2}
\\
\nonumber \\
&& G = -\epsilon r\tilde{r}\frac{f^\prime}{h_\chi}  \left [ \,  4 f^2 \cos^2\theta + (\tilde{r} f^\prime)^2 \sin^2\theta \, \right ]^{-1/2}.
\label{dxdth2}
\eear
Dividing equation (\ref{dxdr2}) by (\ref{dxdth2}) eliminates the dependence on $\theta$, and gives us a first-order ordinary differential equation for $G$.  
The formal solution of this equation is
\be
G = \exp \left ( -\frac{2}{\epsilon}\int d\tilde{r}  \frac{f}{\tilde{r}^2 f^\prime}  \right ),
\ee
which can be put back into equation (\ref{xansatz}), leaving the final coordinate transformation:
\be
 \chi (r,\theta) = \exp \left (-\frac{2}{\epsilon}\int d\tilde{r}  \frac{f}{\tilde{r}^2 f^\prime}  \right ) \cos\theta.
\label{chi2}
\ee
Inserting this back into equation (\ref{eq1}) for the metric coefficient, $h_\chi$, yields
\be h_\chi (r,\theta) = -\epsilon r\tilde{r} f^\prime \exp \left (\frac{2}{\epsilon}\int d\tilde{r}  \frac{f}{\tilde{r}^2 f^\prime}  \right )
 \left [ \,  4 f^2 \cos^2\theta + (\tilde{r} f^\prime)^2 \sin^2\theta \, \right ]^{-1/2}.
\ee
For a given dipole field, one prescribes $f(r)$, and can derive the full coordinate transformation between spherical polar coordinates 
and magnetic coordinates using equations (\ref{stream3}) and (\ref{chi2}).  

As an example of this  we consider a simple power law, $f(\tilde{r})=\tilde{r}^a$, which includes a uniform field ($a=2$) and 
the `standard' dipole field ($a=4$) as special cases.  In this case, the coordinate transformation becomes
\be
\psi(r,\theta) = A\tilde{r}^a\sin^2\theta,\qquad\chi(r,\theta)=C\tilde{r}^{-2/a\epsilon}\cos\theta,
\ee
where $C$ is a constant of integration. 

%%%%%%%%%%%%%%%%%%%%%%%%%%%%%%%%%%%%%%%%%%%%%%%%%%%%%
%%%%%%%%%%%%%%%%%%%%%%%%%%%%%%%%%%%%%%%%%%%%%%%%%%%%%

\section{Some geometric applications of magnetic coordinates}
\label{sec:geom}

As a first application of our magnetic coordinates, we calculate the infinitesimal surface element $dS$ on a sphere of 
radius $r_0$.  This is given by
\be
dS = h_\varphi d\varphi \left (\, h_\psi^2 d\psi^2 + h_\chi^2 d\chi^2  \, \right )^{1/2}.
\ee
On the spherical surface we have $d\chi = -(\partial_\psi r_0/ \partial_\chi r_0) d\psi$ and then
\be
dS = h_\varphi h_\psi \left [\, 1 +  \left ( \frac{h_\chi \partial_\psi r}{h_\psi \partial_\chi r} \right )^2 \, \right ]^{1/2}d\varphi d\psi.
\ee

As a second application of the magnetic coordinates, we derive the equation for a circle of radius $r_0=r_0(\chi, \psi)$.  
Beginning with the stream function, $\psi = A f(\tilde{r}) \sin^2 \theta$, we use equation (\ref{chi2}) to eliminate the $\sin^2\theta$ term:
\be
\frac{\psi}{A} = f \left (1- \frac{\chi^2}{G^2} \right ),
\ee
which is an implicit equation for the circle since $G=G(r_0)$ and $f = f(r_0/R)$.  From this we obtain the partial derivatives 
\be
\frac{\partial r_0}{\partial \psi} = \frac{1}{A} \left [\, \frac{f^\prime}{R} \left ( 1-\frac{\chi^2}{G^2} \right ) + 2f\chi^2\frac{G^\prime}{G^3} \, \right ]^{-1},
\qquad
\frac{\partial r_0}{\partial \chi} = \frac{2f\chi}{G^2} \left [\, \frac{f^\prime}{R} \left ( 1-\frac{\chi^2}{G^2} \right ) + 2f\chi^2\frac{G^\prime}{G^3} \, \right ]^{-1}.
\ee
Their ratio is
\be
\frac{\partial_\psi r_0}{\partial_\chi r_0} = \frac{G}{2Af\cos\theta},
\ee
and we also have
\be
\frac{h_\chi}{h_\psi} = -A \sin\theta \frac{\epsilon\tilde{r}_0 f^\prime}{G}. 
\ee
From these results we find
\be
\frac{h_\chi}{h_\psi} \frac{\partial_\psi r_0}{\partial_\chi r_0} = -\frac{\epsilon}{2} \tan \theta \frac{\tilde{r} f^\prime}{f}.
\ee

In order to make contact with the main body of the paper we also need to calculate the inner product
\be
\hat{r} \cdot \hat{\chi} = \frac{B_0^r}{B_\rP} = 2f \cos\theta [4f^2 \cos^2\theta + (\tilde{r} f^\prime)^2 \sin^2\theta ]^{-1/2}.
\ee
We also have
\be
\hat{r} \cdot \hat{\psi} = \pm [1- (\hat{r} \cdot \hat{\chi})^2 ]^{1/2} = \pm \tilde{r} f^\prime \sin\theta [4f^2 \cos^2\theta + (\tilde{r} f^\prime)^2 \sin^2\theta ]^{-1/2}.
\ee
It is then straightforward to verify the identities:
\be
(\hat{r} \cdot \hat{\chi})  \left [ 1 + \left ( \frac{h_\chi\partial_\psi r_0}{h_\psi \partial_\chi r_0} \right )^2 \, \right ]^{1/2}  = 1,
\qquad 
(\hat{r} \cdot \hat{\psi})  \left [ 1 + \left ( \frac{h_\chi\partial_\psi r_0}{h_\psi \partial_\chi r_0} \right )^2 \, \right ]^{1/2}  =
\pm \frac{1}{2} \tan \theta \frac{\tilde{r} f^\prime}{f}.
\label{id1} 
\ee
For the example given in Appendix~\ref{sec:coordinates}, $f=\tilde{r}^a$, the above formulae reduce to
\be
\frac{\partial_\psi r_0}{\partial_\chi r_0}= -\frac{Ra}{2AC\cos\theta} \left(\frac{r_0}{R}\right)^{-a-2/a\epsilon},
\qquad \left ( \frac{h_\chi\partial_\psi r_0}{h_\psi \partial_\chi r_0} \right )^2 = \frac{a^2}{4} \tan^2\theta,
\qquad \hat{r} \cdot \hat{\chi} = 2\cos\theta \left [\, 4 \cos^2 \theta + a^2 \sin^2\theta \,\right ]^{-1/2},
\ee
and from these the identity (\ref{id1}) follows directly.

%%%%%%%%%%%%%%%%%%%%%%%%%%%%%%%%%%%%%%%%%%%%%%%%
%%%%%%%%%%%%%%%%%%%%%%%%%%%%%%%%%%%%%%%%%%%%%%%

\section{The poloidal Euler equations}
\label{sec:Epol}

The poloidal component of the perturbed Euler equation (\ref{eulerp}) is (after omitting the Coriolis term and the unimportant
$\sim \cB^\prime$ mutual friction term):
\be
4\pi \rho_\rp \nabla_P  \Psi_p = -\nabla_P \left ( \mathbf{B}_0 \cdot \mathbf{b}\right ) + 
\left [\, (\mathbf{b} \cdot \nabla ) \mathbf{B}_0 +   (\mathbf{B}_0 \cdot \nabla ) \mathbf{b}  \, \right ]_P.
\label{eulerP1}
\ee
Using magnetic coordinates, we have for the individual terms of (\ref{eulerP1})
\bear
&& [(\mathbf{b} \cdot \nabla ) \mathbf{B}_0 ]_P = \hat{\psi} \left [ \frac{B_P}{h_\psi h_\chi} \left ( b_\psi \partial_\chi h_\psi - b_\chi \partial_\psi h_\psi \right ) 
-\frac{B^\varphi b^\varphi}{h_\psi h_\varphi}
\partial_\psi h_\varphi  \right ]
+ \hat{\chi} \left [ \mathbf{b} \cdot \nabla B_P - \frac{B^\varphi b^\varphi}{h_\chi h_\varphi}
\partial_\chi h_\varphi  \right ],
\\
\nonumber \\
&& [(\mathbf{B}_0  \cdot \nabla ) \mathbf{b} ]_P = \hat{\psi} \left [ \, \mathbf{B}_0 \cdot \nabla b_\psi - \frac{B_P b_\chi}{h_\psi h_\chi} \partial_\psi h_\chi  
- \frac{B^\varphi b^\varphi}{h_\psi h_\varphi} \partial_\psi h_\varphi \, \right ]
+ \hat{\chi} \left [  \, \mathbf{B}_0 \cdot \nabla b_\chi + \frac{B_P b_\psi}{h_\psi h_\chi} \partial_\psi h_\chi  
- \frac{B^\varphi b^\varphi}{h_\chi h_\varphi} \partial_\chi h_\varphi \, \right ],
\eear
and
\be
\nabla_P \left ( \mathbf{B}_0 \cdot \mathbf{b}\right ) = \frac{\hat{\psi}}{h_\psi} \partial_\psi \left (  B^\varphi b^\varphi 
+ \frac{\partial_\psi \beta}{h^2_\psi h^2_\varphi} \right ) + \frac{\hat{\chi}}{h_\chi} \partial_\chi \left ( B^\varphi b^\varphi 
+ \frac{\partial_\psi \beta}{h^2_\psi h^2_\varphi} \right ).
\ee
Recalling the expressions,
\be
\cT = \varpi b^\varphi, \quad B_\varphi^0 = \cI \nabla\varphi, \quad \mathbf{b}_\rP = \nabla \beta \times \nabla \varphi,
\ee
then after some straightforward manipulations we arrive at the following components of (\ref{eulerP1})
 \bear
&& - 4\pi \rho_\rp \partial_\psi \Psi_\rp =  \partial_\psi \left ( \frac{T \cI}{h_\varphi^2} + \frac{\partial_\psi \beta}{h_\psi^2 h_\varphi^2} \right )
+ \frac{1}{ (h_\chi h_\varphi)^2} \left (\, \partial^2_\chi \beta - K_\chi \partial_\chi \beta  + 2 T \cI \frac{h^2_\chi}{h_\varphi }\partial_\psi h_\varphi 
+ 2\frac{h_\chi}{h_\psi^2} \partial_\psi h_\psi \partial_\psi \beta \, \right ),
\\
\nonumber \\
&& - 4\pi \rho_\rp \partial_\chi\Psi_\rp =  \partial_\chi \left ( \frac{T \cI}{h_\varphi^2} + \frac{\partial_\psi \beta}{h_\psi^2 h_\varphi^2} \right )
- \frac{1}{(h_\psi h_\varphi)^2} \left (\, \partial_\chi \partial_\psi \beta + K_\psi \partial_\chi \beta  
- 2 T \cI \frac{h^2_\psi}{h_\varphi }\partial_\chi h_\varphi \, \right ),
\label{master_pol}
\eear
where
\be
K_\chi = \frac{\partial_\chi h_\varphi}{h_\varphi} + \frac{\partial_\chi h_\chi}{h_\chi} -\frac{\partial_\chi h_\psi}{h_\psi}, \qquad
K_\psi = \frac{\partial_\psi h_\varphi}{h_\varphi} + \frac{\partial_\psi h_\psi}{h_\psi} -\frac{\partial_\psi h_\chi}{h_\chi}.
\ee

%%%%%%%%%%%%%%%%%%%%%%%%%%%%%%%%%%%%%%%%%%%%%%%%
%%%%%%%%%%%%%%%%%%%%%%%%%%%%%%%%%%%%%%%%%%%%


\begin{thebibliography}{}

\bibitem[\protect\citeauthoryear{Abney \& Epstein}{{Abney \& Epstein}}{1996}]{abney}
Abney M., Epstein R.I., 1996, J. Fluid. Mech., {\bf 312}, 327

\bibitem[\protect\citeauthoryear{Akg\"un et al.}{Akg\"un et al.}{2013}]{akgun13}
Akg{\"u}n T., Reisenegger A., Mastrano A., Marchant P., 2013, MNRAS, 433, 2445

\bibitem[\protect\citeauthoryear{Alpar, Langer \& Sauls}{{Alpar, Langer \& Sauls}}{1984}]{als84}
Alpar M. A., Langer S.~A., Sauls J.~A., 1984, ApJ, 282, 533

\bibitem[\protect\citeauthoryear{Andersson et al.}{{Andersson et al.}}{2006}]{NA06}
Andersson N., Sidery T., Comer G.~L., 2006, MNRAS, 368, 162

\bibitem[\protect\citeauthoryear{Braithwaite \& Spruit}{2006}]{braithwaite06a}
Braithwaite J., Spruit H.~C., 2006, A\&A, 450, 1097

\bibitem[\protect\citeauthoryear{Braithwaite \& Nordlund}{2006}]{braithwaite06b}
Braithwaite, J., Nordlund A., 2006, A\&A, 450, 1077

\bibitem[\protect\citeauthoryear{Braithwaite}{2006}]{braithwaite06c}
Braithwaite J., 2006, A\&A, 453, 687

\bibitem[\protect\citeauthoryear{Braithwaite}{2007}]{braithwaite07}
Braithwaite J., 2007, A\&A, 469, 275

\bibitem[\protect\citeauthoryear{Chandrasekhar}{{Chandrasekhar}}{1981}]{chandra}
Chandrasekhar S., 1981, Hydrodynamic and Hydromagnetic Stability, Dover Press, New York

\bibitem[\protect\citeauthoryear{Ciolfi et al.}{2009}]{ciolfi09}
Ciolfi R., Ferrari V., Gualtieri L., Pons J.~A., 2009, MNRAS, 397, 913

\bibitem[\protect\citeauthoryear{Ciolfi \& Rezzolla}{2013}]{CR13}
Ciolfi R., Rezzolla L., 2013, MNRAS, 435, L43
 

\bibitem[\protect\citeauthoryear{Colaiuda et al.}{2008}]{colaiuda08}
Colaiuda  A., Ferrari V., Gualtieri L., Pons J.~A., 2008, MNRAS, 385, 2080

\bibitem[\protect\citeauthoryear{Easson}{{Easson}}{1979}]{easson79}
Easson I., 1979, ApJ, 233, 711

\bibitem[\protect\citeauthoryear{Ferraro}{1954}]{ferraro54}
Ferraro  V.~C.~A., 1954, ApJ, 119, 407

\bibitem[\protect\citeauthoryear{Flowers \& Ruderman}{1977}]{FR77}
Flowers E., Ruderman M.~A., 1977, ApJ, 215, 302

\bibitem[\protect\citeauthoryear{Fujisawa \& Kisaka}{2014}]{fujisawa14}
Fujisawa K., Kisaka S., 2014, MNRAS, 445, 2777

\bibitem[\protect\citeauthoryear{Glampedakis et al.}{2011}]{supercon}
Glampedakis K., Andersson N., Samuelsson L., 2011, MNRAS, 410, 805

\bibitem[\protect\citeauthoryear{Glampedakis et al.}{2012}]{GAL}
Glampedakis K., Andersson N., Lander S.~K.,  2012, MNRAS, 420, 1263

\bibitem[\protect\citeauthoryear{Greenspan}{Greenspan}{1968}]{Greenspan} 
Greenspan H.~P., 1968, The Theory of Rotating Fluids,  Cambridge University Press, Cambridge

\bibitem[\protect\citeauthoryear{Hall \& Vinen}{1956a}]{hall56a}
{Hall}  H.~E., {Vinen} W.~F., 1956, Proc. R. Soc,  London A, 238, 204

\bibitem[\protect\citeauthoryear{Hall \& Vinen}{1956b}]{hall56b}
{Hall} H.~E., {Vinen} W.~F., 1956, Proc. R. Soc.  London A, 238, 215

\bibitem[\protect\citeauthoryear{Ho et al.}{2012}]{HGA12}
Ho W.~C.~G., Glampedakis K., Andersson N., 2012, MNRAS, 422, 2632

\bibitem[{{Kaminker} {et al.}(2006)}]{kaminkeretal06}
Kaminker A.~D., Yakovlev D.~G., Potekhin A.~Y., Shibazaki N., Shternin P.~S., Gnedin O.~Y.
2006, MNRAS, 371, 477


\bibitem[\protect\citeauthoryear{Lander \& Jones}{{Lander \& Jones}}{2012}]{lander12}
Lander S.~K., Jones D.~I.,  2012, MNRAS, 424, 482 

\bibitem[\protect\citeauthoryear{Lasky et al.}{2011}]{lasky11}
Lasky P.~D., Zink B., Kokkotas K.~D., Glampedakis K., 2011, ApJ, 735, L20


\bibitem[\protect\citeauthoryear{Lasky \& Melatos}{2013}]{lasky13}
Lasky P.~D., Melatos, A. 2013, Phys. Rev. D, 88, 103005


\bibitem[\protect\citeauthoryear{Mastrano \& Melatos}{2005}]{mastrano05}
Mastrano A., Melatos  A., 2005, MNRAS, 361, 927

\bibitem[\protect\citeauthoryear{Mastrano et al.}{{Mastrano et al.}}{2011}]{mastrano11}
Mastrano A., Melatos  A., Reisenegger A., Akg\"un T., 2011, MNRAS, 417, 2288


\bibitem[\protect\citeauthoryear{Melatos}{{Melatos}}{2012}]{melatos12}
Melatos  A., 2012, ApJ, 761, 32


\bibitem[\protect\citeauthoryear{Mendell}{{Mendell}}{1998}]{mendell98}
Mendell G., 1998, MNRAS, 296, 903

\bibitem[\protect\citeauthoryear{Mitchell et al.}{{Mitchell et al.}}{2015}]{mitchell15}
Mitchell, J.~P., Braithwaite, J., Reisenegger, A., Spruit, H., Valdivia, J.~A., Langer, N., 2015, MNRAS, 447, 1213

\bibitem[\protect\citeauthoryear{Pons \& Geppert}{{Pons \& Geppert}}{2007}]{pons07}
Pons J.~A., Geppert U., 2007, A\&A , 470, 303

\bibitem[{{Page} {et~al.}(2011)}]{pageetal11}
Page D., Prakash M., Lattimer J.~M., Steiner A.~W., 2011, Phys. Rev. Lett., 106, 081101

\bibitem[{{Shapiro} \& {Teukolsky}(1983)}]{STbook}
{Shapiro} S.~L., {Teukolsky} S.~A., 1983, Black Holes, White Dwarfs, and Neutron Stars.
Wiley, New York


\bibitem[{{Shternin} {et al.}(2011)}]{shterninetal11}
Shternin P.~S., Yakovlev D.~G., Heinke C.~O., Ho W.~C.~G., Patnaude D.~J., 2011, MNRAS, 412, L108

\bibitem[\protect\citeauthoryear{Tayler}{{Tayler}}{1973}]{tayler73}
Tayler  R.~J., 1973, MNRAS, 161, 365

\bibitem[\protect\citeauthoryear{Thompson \& Duncan}{{Thompson \& Duncan}}{2001}]{TD01}
Thompson C., Duncan R.~C., 2001, ApJ, 561, 980

\bibitem[\protect\citeauthoryear{Wright}{1973}]{wright73}
Wright G.~A.~E., 1973, MNRAS, 162, 339


\end{thebibliography}
\end{document}